\pdfoutput=1

\documentclass[11pt]{article}

\usepackage[]{acl}

\usepackage{times}
\usepackage{latexsym}
\usepackage{hyperref}
\usepackage{url}
\usepackage{amsmath}
\usepackage{tcolorbox}
\usepackage{alphabeta}
\usepackage{fancyvrb}
\usepackage{booktabs} 
\usepackage{graphicx}
\usepackage{multirow}

\usepackage{color, colortbl}
\definecolor{beige}{rgb}{0.98, 0.94, 0.9}

\usepackage[nameinlink,noabbrev,capitalise,sort]{cleveref}
\usepackage{subcaption}
\usepackage{enumitem}

\usepackage[T1]{fontenc}

\usepackage[utf8]{inputenc}

\usepackage{microtype}

\usepackage{inconsolata}

\usepackage{graphicx}
\definecolor{lightmagenta}{rgb}{0.95, 0.8, 1.0}

\usepackage{pifont}
\newcommand{\cmark}{\ding{51}}%
\newcommand{\xmark}{\ding{55}}%

\newcommand{\imagecaptioning}{\raisebox{-.2ex}{\includegraphics[height=2ex]{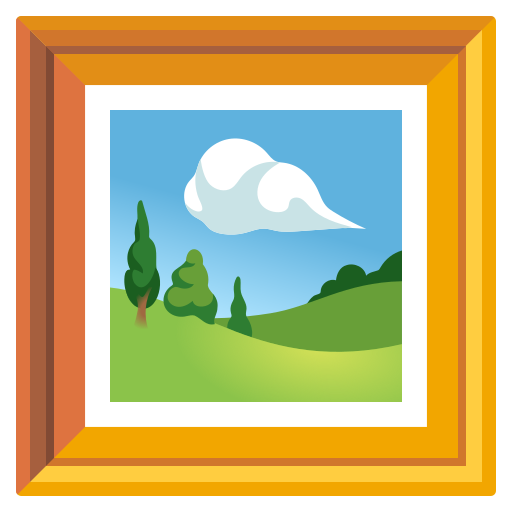}}}
\newcommand{\usersurvey}{ \raisebox{-.2ex}{\includegraphics[height=2ex]{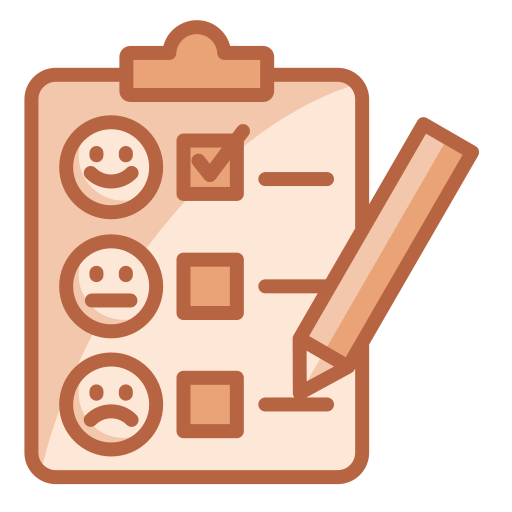}}}
\newcommand{\imageqa}{\raisebox{-.2ex}{\includegraphics[height=2ex]{images/image_des.png}} }
\newcommand{\videoOR}{{\includegraphics[height=2ex]{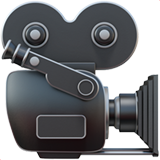}}}
\newcommand{\Braille}{{\includegraphics[height=2.5ex]{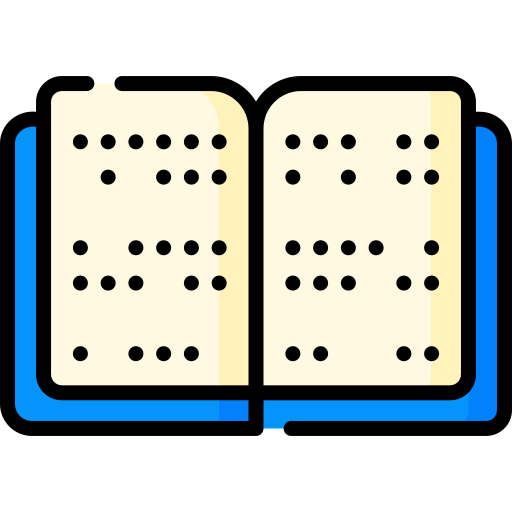}}}
\newcommand{\videoqa}{\raisebox{-.2ex}{\includegraphics[height=2ex]{images/video.png}}}

\title{Evaluating Multimodal Language Models as Visual Assistants for Visually Impaired Users}

\author{Antonia Karamolegkou*$^{\dagger}$ \ \ Malvina Nikandrou*$^{\ddagger}$ \ \ Georgios Pantazopoulos*$^{\ddagger}$ \\
        Danae Sanchez Villegas$^{\dagger}$ \ \ Phillip Rust$^{\dagger}$ \ \ Ruchira Dhar$^{\dagger}$ \ \ Daniel Hershcovich$^{\dagger}$ \ \ Anders S{\o}gaard$^{\dagger}$ \vspace{0.2cm} \\
        $^{\dagger}$ University of Copenhagen \ \ 
        $^{\ddagger}$ Heriot-Watt University \\
        *Equal contribution \\
}

\begin{document}
\maketitle
\begin{abstract}

This paper explores the effectiveness of Multimodal Large Language models (MLLMs) as assistive technologies for visually impaired individuals. We conduct a user survey to identify adoption patterns and key challenges users face with such technologies. Despite a high adoption rate of these models, our findings highlight concerns related to contextual understanding, cultural sensitivity, and complex scene understanding, particularly for individuals who may rely solely on them for visual interpretation. Informed by these results, we collate five user-centred tasks with image and video inputs, including a novel task on Optical Braille Recognition. Our systematic evaluation of twelve MLLMs reveals that further advancements are necessary to overcome limitations related to cultural context, multilingual support, Braille reading comprehension, assistive object recognition, and hallucinations. This work provides critical insights into the future direction of multimodal AI for accessibility, underscoring the need for more inclusive, robust, and trustworthy visual assistance technologies\footnote{We make our survey, evaluation data, and code publicly available at \url{https://github.com/MalvinaNikandrou/visual-assistant-eval}}. 
\end{abstract}

\section{Introduction}

As the capabilities of Large Language Models (LLMs) have been extended to multimodal contexts, particularly in applications that combine vision and language processing, one promising area is the use of multimodal LLMs (MLLMs) as visual assistants.
MLLMs can provide valuable support, particularly for individuals with visual impairments, by accurately interpreting and describing visual content, including real-world images and videos \citep{karamolegkou-etal-2024-vision}. 

\begin{figure}[tb]
    \centering
    \includegraphics[width=0.99\columnwidth]{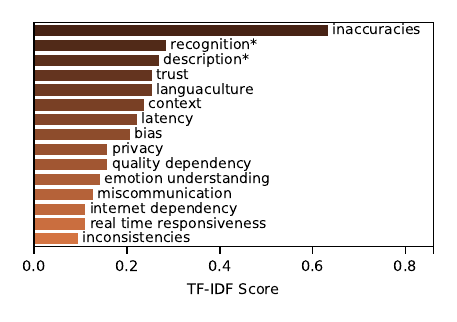}
    \caption{User survey results highlighting the 15 most important terms (measured by TF-IDF scores), representing key challenges for AI visual assistants. (*) includes tasks such as object, handwriting and face recognition; and image, scene, and video description.}
    \label{fig:survey}
\end{figure}

MLLMs have already integrated into assistive technologies and services\footnote{\url{https://aira.io/}, \url{https://www.bemyeyes.com/}}, such as automated captioning systems and smart devices \citep{yuan2025walkvlmaidvisuallyimpairedpeople}.
However, these models still face limitations in acting as effective visual assistants \citep{eyes_shut}.
For example, a Blind or Low Vision (BLV) traveller using an MLLM-powered assistant to navigate a foreign city may receive inaccurate descriptions of street signs due to poor image quality or incomplete translations, leading to confusion or safety risks. Such scenarios raise concerns about reliability and safety in critical tasks and pose risks for users who depend on precise visual assistance. Given that modern MLLMs are hill-climbing multimodal reasoning benchmarks \citep{liu2024mmbench, li2024mvbench, wang2024measuring}, a comprehensive evaluation of their effectiveness and limitations in accessibility applications is urgently needed.

\begin{figure*}[tb]
    \centering
    \includegraphics[width=\textwidth]{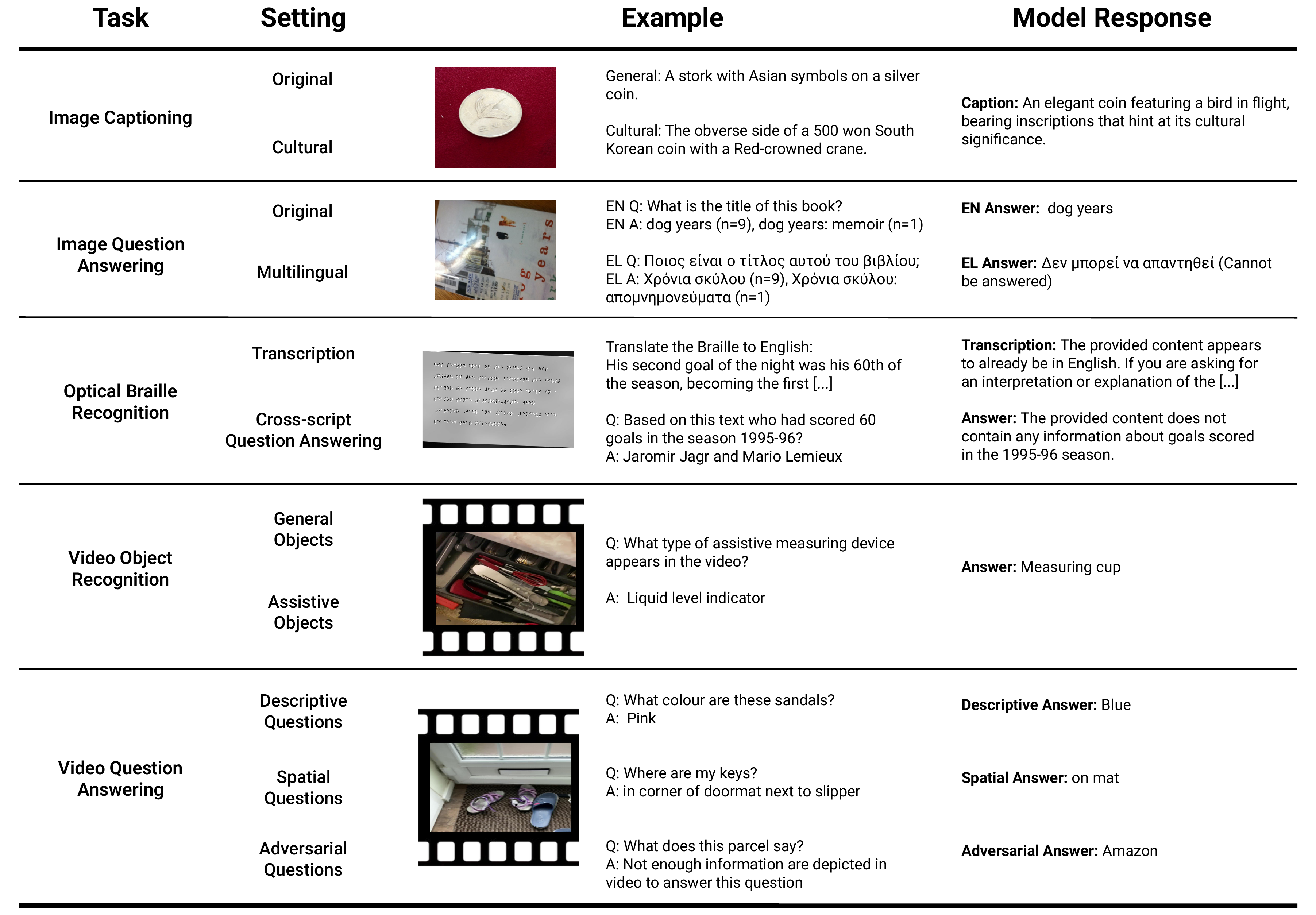}
    \caption{Illustration of the five key areas of our framework. We focus on tasks pertinent to BLV people covering different aspects for captioning, transcribing, and answering questions about visual content.}
    \label{fig:main_fig}
\end{figure*}

To better understand these challenges, we conducted a user survey \usersurvey\hspace{0.01cm}  (\S \ref{sec:usersurvey}) to identify the tasks and issues most relevant to individuals with visual impairments. 
\cref{fig:survey} summarizes the survey responses, highlighting that inaccuracies, such as hallucinations (i.e., factually incorrect or inconsistent generated content) and misleading information, are the primary concerns. 

Motivated by the findings of our survey, we design an evaluation framework with tasks relevant to BLV individuals, focusing on five key areas as shown in \cref{fig:main_fig}: 1) Image Captioning \imagecaptioning ~targeting cultural content, 2) multilingual Image Question Answering \imageqa, 3) Optical Braille Recognition \Braille ~to transcribe and answer questions about Braille text rendered in images, 4) Video Object Recognition \videoOR ~covering general usage objects as well as assistive items commonly used by BLV people, and 5) Video Question Answering \videoqa{} ~covering descriptive, spatial, and adversarial questions. 
Importantly, we contribute datasets for multilingual and video question answering as well as Braille recognition to improve the capabilities of the next generation of MLLMs that assist BLV individuals. 
Our experiments emphasize the need for further advances in multimodal AI to ensure these models can reliably support individuals who rely on them for visual tasks.

\section{Related Work}
\label{sec:background}

\paragraph{MLLM Evaluation Benchmarks} 

MLLMs are mainly evaluated on general-domain benchmarks that assess visual perception, knowledge, and reasoning \citep{goyal2017making, schwenk2022okvqa, yin2023survey, Li_2024_CVPR, liu2024mmbench, lu2024mathvista}. However, these benchmarks do not capture all critical dimensions of MLLM performance. One exception is the holistic evaluation by \citet{lee2024vhelm}, which examined  22 MLLMs across nine aspects, revealing that no model excels in all areas and that all lack multilingual support. Other studies highlight inconsistencies in MLLMs' responses \citep{chen-etal-2024-measuring} and measure performance in diverse cultural contexts \citep{nayak-etal-2024-benchmarking, mogrovejo2024cvqa}. Despite these efforts, the effectiveness of MLLMs as visual assistants in accessibility settings remains unexplored.

\paragraph{Multimodal Models for Assistive Applications} Previous works apply \textit{task-specific} models for assistive applications including visual question answering \citep{liu2024right, huh2024longform, gurari2018vizwiz}, image captioning \citep{gurari2020captioning}, object detection \citep{reynolds2024salient, tseng2022vizwiz}, and private content identification \citep{tseng2024biv}. Some conversational agents focus on privacy-aware assistance \citep{baker-etal-2021-spoon}, education for BLV users \citep{di-nuovo-etal-2024-educational}, scenarios with low-quality images \citep{yang2024viassist}, while other studies integrate MLLMs into assistive devices and smartphone applications \citep{holiel2024assisting, de2024vqask}. These works demonstrate the potential of MLLMs in accessibility but also highlight the need for systematic evaluation of their effectiveness and limitations.

\section{User Survey \usersurvey}
\label{sec:usersurvey}
Understanding user perspectives is crucial for identifying key application areas, surfacing unmet needs, and guiding future improvements in model design and evaluation \citep{researchtools, kirk2024the}.
To gain insights into the real-world usability of MLLMs in the role of visual assistants, we conducted a survey focused on user adoption patterns and experienced challenges.

\paragraph{Design}
The survey consists of two phases: open-ended questions and Likert scale ratings. Phase one begins with a user adoption question, asking whether participants use or would consider using AI for visual assistance. The second question explores tasks where these models could be most beneficial, while the third targets challenges users have experienced in past interactions. Phase two focuses on assessing specific tasks and issues. We recruited $106$ participants through Prolific\footnote{\url{https://www.prolific.com/}}, with varying degrees of visual impairment. By analysing responses and identifying key themes, we identified common use cases and areas where AI needs improvement. Below, we present a summary of the open-ended responses and provide further details on survey design, analysis, compensation, demographics, and Likert scale results in \cref{appendix:survey}.

\paragraph{User Adoption and Tasks}
The majority of respondents (87\%) use or would use AI as visual assistants, while 9\% declined due to concerns about accuracy, reliability, and lack of personal touch, and 4\% were unsure, depending on the assistance type. Participants found AI most useful for description, transcription, translation, and recognition. Common use cases included identifying and translating products for shopping or cooking, understanding diagrams in subjects like chemistry and math, analysing food consistency, choosing outfits, and interpreting facial expressions. Some mentioned more specialized uses, such as autonomous navigation, medical imaging analysis, Braille interpretation, space planning, design assistance, artistic creation, and emotional support.

\paragraph{Challenges}
Participants were asked to list challenges they have experienced when using AI models. Responses varied in specificity, requiring grouping and qualitative analysis using an iterative open thematic approach\footnote{Further details are in \cref{appendix:survey}.} \citep{researchtools}.
The most common challenges are visualized in \cref{fig:survey}. A major problem was \emph{inaccuracies}, as it was often mentioned that users struggle to verify whether the provided information is correct. This included issues such as \textit{incorrect directions}, \textit{misidentification of objects}, \textit{misinterpretation of signs and symbols}, and \textit{misleading or incomplete responses}. Many challenges fell under \emph{recognition} and \emph{description} tasks, particularly difficulties with handwritten text (especially small, messy, or multilingual), Braille, currency, and signs. Participants also reported problems with object recognition in poor image conditions (low resolution, lighting issues, or background noise) and in cluttered or ambiguous settings, sometimes mistaking shadows for obstacles or misidentifying overlapping objects. 

Some also mentioned that AI often fails to recognize hazards, interpret multicultural and social cues, and provide sufficiently detailed descriptions. Further challenges involved trust, language limitations, contextual understanding in scenes (e.g., understanding spatial relationships or complex environments), latency, bias, privacy concerns, dependency on high-quality data, emotional understanding, and communication barriers, such as unclear prompts or lack of adaptation to individual needs.

\section{Evaluation Framework}
\label{sec:experiments}
We evaluate MLLMs across image and video understanding tasks, specifically designed to assist visually impaired individuals.
Our task selection is informed by user input, reflecting use cases where users reported a high likelihood of adopting AI visual assistants (see \cref{fig:result-tasks-likert}).
We additionally emphasize high-priority needs such as cultural context awareness, multilingual support, and recognition of assistive devices and hallucinations.

\paragraph{Tasks} 
The evaluation spans five tasks:
\texttt{Image Captioning}, which evaluates performance in generating descriptions for images taken by visually impaired individuals (\S \ref{sec:image-captioning}). \texttt{Image Question Answering} to evaluate question answering using images and questions provided by visually impaired individuals (\S \ref{sec:imageqa}). \texttt{Optical Braille Recognition}, where we assess the performance on transcribing and answering questions about Braille text in images (\S \ref{sec:Braille}).
Finally, using videos recorded by BLV individuals, we evaluate \texttt{Video Object Recognition} (\S \ref{sec:videoobj}), and \texttt{Video Question Answering} on descriptive, spatial, and adversarial questions (\S \ref{sec:videoqa}). In each corresponding section, we introduce the related background, specify the evaluation setup, and report our result.

\paragraph{Models} We focus on 12 prominent models based on the following criteria: 1) strong performance in image and video understanding, 2) open access (open-source or open-weights), and 3) moderate computational overhead to balance performance and latency. \cref{tab:models} provides details about multilingual support and exposure to domain data.

\section{Image Captioning {\includegraphics[height=2ex]{images/image_des.png}}}
\label{sec:image-captioning}

Image captioning aims at generating textual descriptions for images.  \citet{gurari2020captioning} introduce VizWiz, the first dataset with images from visually impaired users, launching a series of multimodal challenges \citep{gurari2018vizwiz}. Since then, research has focused on improving models for assisting visually impaired users \citep{lessons_learned, ahsan-etal-2021-multi, delloul2023real}, mostly in English settings. More recently, \citet{karamolegkou-etal-2024-vision} identified cultural implicatures in VizWiz that annotations and models overlook and curated a subset of 324 images and 648 captions spanning 60 cultures.

\paragraph{Setup} We evaluate model performance on generating descriptions for images taken by visually impaired people.
We use the original validation set of VizWiz-Captions (N=$500$) \cite{gurari2020captioning} and the \textbf{multicultural} extension \cite{karamolegkou-etal-2024-vision} (N=$324$), which filters the original set and provides re-annotations focused on culture-related content.
As a metric, we use the RefCLIPScore \cite{hessel2021clipscore}, which has shown robust alignment with human judgment.

\begin{table}[tb]
    \centering
    \small
    \begin{tabular}{lcc}\toprule
    \textbf{Model}    &  \textbf{Original} & \textbf{Cultural} \\\midrule
    Idefics3   & 76.0 & 75.5 \\
    InternVL2.5-MPO & 74.3 & 74.8 \\
    LLaVA-v1.6   & 72.3 & 52.2 \\
    Llama-3.2-Vision-Instruct   & 75.0 & 72.8 \\
    MiniCPM-V-2.6  & 78.0 & 74.8  \\
    Molmo   & 70.9 & 47.4 \\
    PaliGemma   & \textbf{81.0} & 55.0 \\
    Phi-3.5-Vision-Instruct   & 71.9 & 62.6 \\
    Qwen2-VL-Instruct & 75.9 & \textbf{76.9} \\
    \bottomrule
    \end{tabular}
    \caption{RefCLIPScore results on the original and cultural VizWiz image captioning validation set.}
    \label{tab:vizwiz_cap}
\end{table}

\paragraph{Results} \cref{tab:vizwiz_cap} shows the image captioning evaluation results. 
All models achieve RefCLIPScores between 70 and 81 on the original setting indicating relatively good performance. PaliGemma outperforms other MLLMs by at least 5 points, likely due to its exposure to VizWiz data during pretraining. In the cultural setting,  we observe a clear performance divide. Five out of the examined MLLMs show robust performance ($\pm$4 points difference), while other models show substantial degradation (20-25 points). To assess progress in culture-aware descriptions, we inspect 100 captions from the top two models. For both Qwen2-VL-Instruct and Idefics3, approximately a third of the generated captions (31\% and 33\%, respectively) include accurate but generic information--while they correctly describe the scene, they miss culturally significant details such as specific names of symbols, cultural figures, or non-English language scripts. This indicates that models might still overlook cultural context, which is essential to fully describe a scene.

\section{Image Question Answering {\includegraphics[height=2ex]{images/image_des.png}}}
\label{sec:imageqa}
\begin{table}[tb]
    \centering
    \small
    \begin{tabular}{lcc}\toprule
    \textbf{Model} & \textbf{Original} & \textbf{Multilingual} \\\midrule
    Idefics3 & 45.7 &  30.4 \\
    InternVL2.5 & 65.1 & 39.1\\
    LLaVA-v1.6 & 54.8 & 40.8\\
    Llama-3.2-Vision-Instruct & 52.9 & 29.6 \\
    MiniCPM-V-2.6 & 72.2 & 30.7 \\
    Molmo & 40.2 & 28.6 \\
    PaliGemma   & \textbf{75.6} & 16.9 \\
    Phi-3.5-Vision-Instruct & 59.0   &  36.5 \\
    Qwen2-VL-Instruct & 61.9 & \textbf{44.9} \\
    \bottomrule
    \end{tabular}
    \caption{VQA Accuracy results on the original and multilingual VizWiz question answering validation set.}
    \label{tab:vizwiz_vqa}
\end{table}
Image question answering (IQA) enables users to ask about images and receive relevant answers. As part of the VizWiz initiative, \citet{gurari2018vizwiz} created an IQA dataset capturing real-world challenges, where visually impaired users take photos that may be blurry, poorly framed, or contain unanswerable questions. Recent efforts address these issues through answer grounding \citep{10377187}, long-form answers \citep{huh2024longform}, and models suggesting image adjustments \citep{liu2024right}. However, no work has yet examined these challenges in a multilingual setting.

\paragraph{Setup}
We evaluate each model on visual question answering using the VizWiz validation set \cite{gurari2018vizwiz}.
To assess the global accessibility of these models, we extend the evaluation to a \textbf{multilingual} setting.
We use an automatic translation pipeline with human quality checks to translate 500 questions and reference answers to 34 languages. Details about the translation process are provided in \cref{sec:multi}.
The task metric is VQA accuracy \cite{VQA}, which takes into account multiple reference answers as the evaluation metric.

\paragraph{Results}
The results shown in \cref{tab:vizwiz_vqa} reveal large performance disparities across models and evaluation settings.
In the original English setting, PaliGemma and MiniCPM-V-2.6 (75.6\% and 72.2\% respectively), which include VizWiz VQA data in their pretraining mixture, achieve the highest performance by a large margin. 
However, these models also suffer the largest performance drops in the multilingual setting.
We observe that they often fail to follow prompt instructions, such as answering in the language of the question, instead defaulting to English responses.
The best multilingual performance is achieved by Qwen2-VL-Instruct, which shows the most consistent performance ranging between 35.4 and 49.0 for all non-English languages. \cref{tab:vizwiz_vqa_resource} shows the VQA accuracy grouped by high-, medium-, and low-resourced languages \cite{joshi-etal-2020-state}. We observe limited performance variance across the three groups, with all models performing similarly poorly regardless of language resource levels, suggesting that even high- and medium-resource languages lack reliable IQA support for blind users who do not speak English.

\section{Optical  Braille Recognition {\includegraphics[height=2.6ex]{images/Braille.png}}}
\label{sec:Braille}

Despite increasing interest in the text comprehension abilities of MLLMs \citep{li2024seed2plus, Liu_2024}, their capacity to process Braille within images remains underexplored.

Existing Braille recognition approaches focus on character-level classification where a visual component first detects the characters, followed by a character classifier \citep{li2020optical, smelyakov2018Braille, gao2024enhanced}.
However, character-level approaches do not fully assess the reading comprehension capabilities of modern MLLMs.
For this purpose, we compile two datasets focusing on sentence-level Braille-to-Text transcription and paragraph-level cross-script question-answering.
Our datasets differ from prior work as they target longer context, support zero-shot and few-shot evaluation, and introduce a training split that can be incorporated in the visual instruction tuning data of an MLLM.

\subsection{Dataset Creation}
\label{sec:Braille_dataset_creation}

For sentence-level transcription, we compile a dataset using English sentences from the shared task of WMT 2024 \citep{wmt-2024-1}.
More specifically, we use a subset of 100k sentences from the \texttt{Facebook-wikimatrix-1-deu-eng} corpus for training, as well as NTREX-128 \citep{federmann-etal-2022-ntrex} (N=1997), and FLORES-200 \citep{nllb-24} (N=1012) for evaluation.
With regards to paragraph-level question answering, we leverage SQuAD \citep{rajpurkar-etal-2018-know} (training N=130K, evaluation N=11.9K), which provides text paragraphs together with a few relevant questions.
In both tasks, we render the Braille text into images (see \cref{appendix:obr}), and apply augmentations that correlate with quality flaws often occurring in images taken by BLV people \citep{yu2023quality}.
The model accepts an image containing Braille text, the input prompt including a question for SQuAD only, and needs to provide the appropriate English response, i.e. either the transcription of the rendered Braille sentence or the answer to the question.

\paragraph{Evaluation Metrics} Since the Braille-to-Text transcription is a character-level transformation, for the sentence-level transcription, we report the chrF++ score \citep{popovic-2017-chrf}. 
For SQuAD, we report the character-level F1-score based on the model's prediction and the candidate answers for each question, as well as the exact match.

\begin{table}[tb]
    \centering
    \small
    \begin{tabular}{lccc}\toprule
       \textbf{ Model}    &  \textbf{F-200} &\textbf{ N-128} & \textbf{Avg}\\\midrule
        Idefics3  & 1.9 & 2.1 & 2.0\\
        InternVL2.5-MPO & 8.7 & 8.5 & 8.6\\
        LLaVA-v1.6 & 2.9 & 2.7 & 2.8\\
        Llama-3.2-Vision-Instruct & 8.9 & 8.3 & 8.5\\
        MiniCPM-V-2.6 & 8.9 & 9.2 & 9.1\\
        Molmo & 5.3 & 5.44 & 5.4\\
        Phi-3-vision-128k-instruct & 10.2 & 9.5 & 9.9\\
        Qwen2-VL-Instruct & \textbf{75.2} & \textbf{72.5} & \textbf{73.8}\\
        \bottomrule
    \end{tabular}
    \caption{Zero-shot chrF++ scores on sentence-level Braille-to-Text transcription for FLORES-200 and NTREX-128 datasets. Out of the eight models, only Qwen2-VL-Instruct exhibits Braille comprehension capabilities.}
    \label{tab:obr_zeroshot}
\end{table}

\subsection{Results}
\begin{figure*}[tb]
\minipage{0.47\textwidth}
  \includegraphics[width=\linewidth]{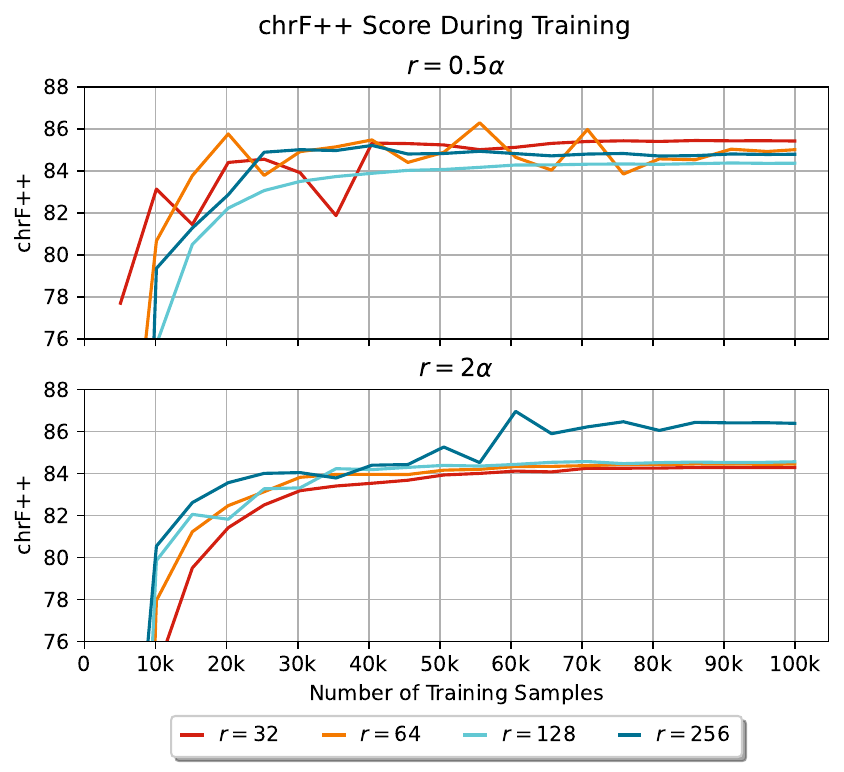}
\endminipage\hfill
\minipage{0.47\textwidth}
  \includegraphics[width=\linewidth]{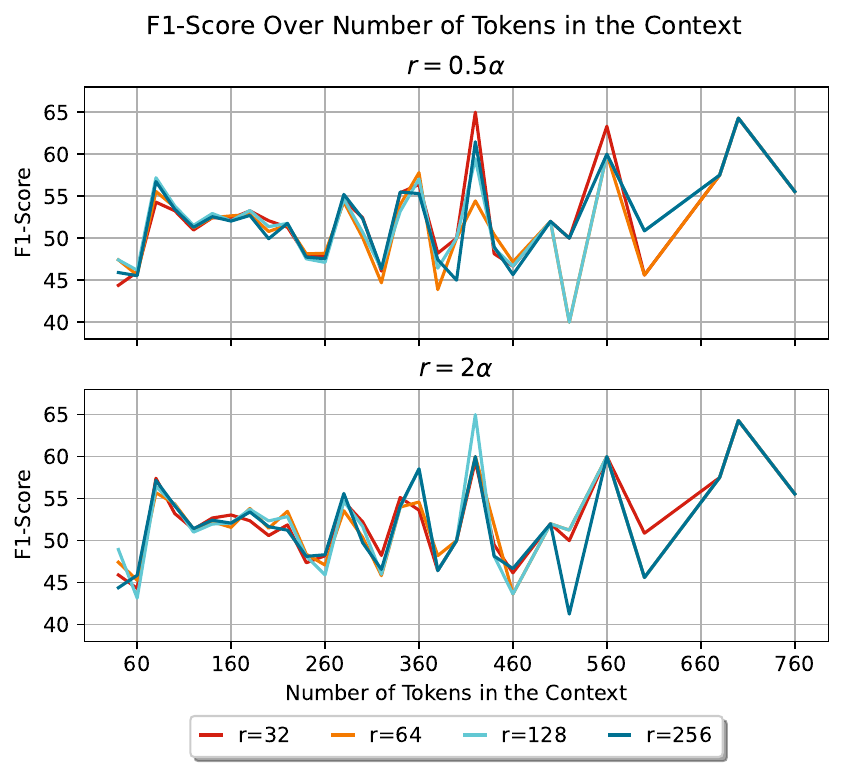}
\endminipage
    \caption{Left: Average chrF++ on sentence-level Braille-to-Text transcription. Right: F1-Score on cross-script question answering where results are binned based on the length of the context paragraph.}
    \label{fig:obr_all}
\end{figure*}
\paragraph{Can MLLMs read Braille?} We prompt MLLMs to transcribe rendered Braille sentences to regular English text. \cref{tab:obr_zeroshot} illustrates the zero-shot performance on our two English-to-Text transcription evaluation sets.
Our results clearly show that most modern MLLMs are not equipped with Braille recognition capabilities. Surprisingly, out of the examined models, only Qwen2-VL-Instruct demonstrates non-trivial performance indicating its capability of reading Braille from images.

\begin{table}[tb]
    \centering
    \small
    \begin{tabular}{cc |ccc| cc}
    \toprule
    & & \textbf{F-200} &\textbf{ N-128} & \textbf{Avg} & \multicolumn{2}{c}{\textbf{SQuAD}}\\
    r & \alpha & \multicolumn{3}{c|}{chrF++} &  F1 & EM \\
    \toprule
    32 & 64 & 88.2 & 82.6 & 85.4 & 51.8 & 49.8\\
    64 & 128 & 88.2 & 81.9 & 85.0 & 51.8 & 49.8\\
    128 & 256 & 87.4 & 81.4 & 84.4 & 52.0 & 50.0\\
    256 & 512 & 87.6 & 81.9 & 84.8 & 51.7 & 49.7\\
    \midrule
    32 & 16 & 87.2 & 81.4 & 84.3 & 51.9 & 49.9\\
    64 & 32 & 87.4 & 81.4 & 84.4 & 51.9 & \textbf{50.2}\\
    128 & 64 & 87.5 & 81.7 & 84.6 & 51.9 & 50.1\\
    256 & 128 & 89.2 & 83.5 & \textbf{86.4} & \textbf{52.1} & 50.1\\
    \bottomrule
    \end{tabular}
    \caption{LoRA fine-tuning results for Llama-3.2-Vision-Instruct on sentence-level Braille-to-Text transcription (FLORES-200, NTREX-128, average) and cross-script question answering (SQuAD).}
    \label{tab:obr_finetune_wmt}
\end{table}

\paragraph{Proof of concept: Learning to Read Braille}
Next, we are interested in a training recipe that results in an MLLM capable of reading Braille text in images.
For this purpose, we focus exclusively on Llama-3.2-Vision-Instruct, as a model with strong text comprehension capabilities but lacking the ability to read Braille text. We finetune Llama-3.2-Vision-Instruct both for sentence-level transcription, as well as paragraph-level question answering using LoRA \citep{hu2022lora} following guidelines from existing cookbook recipes\footnote{\href{https://magazine.sebastianraschka.com/p/practical-tips-for-finetuning-llms}{Practical Tips for Finetuning LLMs Using LoRA}}.
For each configuration, we sweep across different hyperparameters (see \cref{appendix:obr}) and select the one with the best validation performance.

\cref{tab:obr_finetune_wmt} illustrates the finetuning results of Llama-3.2-Vision-Instruct on both tasks. We observe that the model achieves great performance across a wide range of configurations. Additionally, \cref{fig:obr_all} (left) shows that model performance improves quickly--typically saturating at 30K samples. Similar results can be observed in the case of paragraph-level question answering.
Finally, \cref{fig:obr_all} (right) shows the F1-Score of all finetuning runs according to the length of the context paragraph, i.e., the number of English tokens that have been transcribed to Braille and rendered in images. We observe that the model maintains similar performance in short as well as long paragraphs. Taken together, these results show that while most modern MLLMs are not equipped with Braille comprehension, learning to read Braille text in images is feasible with a moderate number of demonstrations. Consequently, we expect the next generation of MLLMs powering accessibility applications to take into account Braille reading comprehension capabilities as part of the visual instruction tuning stage.

\section{Video Object Recognition {\includegraphics[height=2ex]{images/video.png}}}
\label{sec:videoobj}

\begin{table}[tb]
    \centering
    \small
    \begin{tabular}{lcc}\toprule
    \textbf{Model}    & \textbf{General} & \textbf{Assistive} \\
    & (N=880) & (N=156)\\\midrule
 \multicolumn{3}{l}{\textit{Video LMs}}    \\\midrule
   LLaVA-NeXT-Video & 56.0 & 26.0 \\
   LLaVA-Video & 65.7 & 41.3 \\
   VideoChat-Flash & 56.0 & 20.8 \\\midrule
 \multicolumn{3}{l}{\textit{Image + Video LMs}}    \\\midrule
   InternVL2.5-MPO & 59.1 & 36.5 \\
   MiniCPM-V-2.6 & 65.1 & \textbf{44.2} \\
   Phi-3.5-Vision-Instruct & 52.2 & 25.3 \\
   Qwen2-VL-Instruct & \textbf{69.8} & 39.7 \\
    \bottomrule
    \end{tabular}
    \caption{Accuracy in Video Object Recognition of general and assistive object categories.}
    \label{tab:orbit_video_qa}
\end{table} 

Video-based object recognition extends traditional image recognition \cite{ILSVRC15, hu2023open, webly}, allowing models to identify objects that appear in a video sequence. While image recognition provides a snapshot of visual content, it may miss useful contextual cues available in videos, such as gradual occlusions or varying viewpoints. 
Unlike video classification, which typically focuses on activity recognition \cite{goyal2017something, kay2017kinetics}, our task aims to identify the presence of objects in a video, making it more aligned with real-world assistive applications.
Moreover, while there are a number of datasets filmed in real-world environments \citep{core, Damen_2018_ECCV}, very few are specifically centred on visual assistance for visually impaired users \citep{orbit, blv}.

\paragraph{Setup} We evaluate models on their ability to identify objects from videos recorded by BLV people. Similar to the image settings, these videos pose challenges such as blurriness and non-centred objects. We use 1036 video clips from ORBIT \cite{orbit}, which show household objects from 92 categories. These objects include both general everyday objects (e.g., TV remote control) and assistive items (e.g., Braille display). Additionally, objects are recorded in \textit{Clean} videos, which show an object in isolation, and \textit{Clutter} videos, which show the target object in context with other items.
More details about the dataset are provided in \cref{appendix:vor}. Following previous work on evaluating generated outputs with one ground truth answer, we adopt the LAVE protocol \citep{manas2024improving}, which leverages a language model to judge the generated outputs and provide a rating between 1-3. We use Llama-3.3-70B-Instruct \citep{llama3modelcard} and report average normalized ratings.

\paragraph{Results}
\cref{tab:orbit_video_qa} reports model accuracy on recognizing general and assistive objects, revealing a clear gap: while models perform moderately well on generic object categories (52-69.8\% accuracy), they struggle significantly at recognizing assistive items, achieving only 23-41\% accuracy.
This performance disparity might be expected from a data distribution perspective, as assistive objects are less common and current MLLMs are known to struggle with capturing long-tail knowledge \cite{geigle-etal-2024-african, parashar2024neglected}.
However, this result indicates that generalist MLLMs are still far from providing comprehensive support for visually impaired users in everyday contexts.

\section{Video Question Answering {\includegraphics[height=2ex]{images/video.png}}}
\label{sec:videoqa}
\begin{table}[tb]
    \centering
   \small
    \begin{tabular}{lcccc}\toprule
    \textbf{Model}    &  \textbf{D} & \textbf{S} & \textbf{A} & \textbf{Avg} \\\midrule
 \multicolumn{3}{l}{\textit{Video LMs}}    \\\midrule
   LLaVA-NeXT-Video & 56.0 & 49.7 & 13.4 & 39.7  \\
LLaVA-Video  & \textbf{78.2} & 63.4 & 7.7 & 49.8\\
VideoChat-Flash & 72.4 & \textbf{64.1} & 9.2 & 48.6 \\
\midrule
\multicolumn{3}{l}{\textit{Image + Video LMs}}    \\\midrule
InternVL2.5-MPO  & 67.7 & 59.4 & 9.0 & 45.4 \\
MiniCPM-V-2.6  & 68.7 & 63.3 & \textbf{17.7} & \textbf{49.9} \\
Phi-3.5-Vision-Instruct  & 61.4 & 46.3 & 10.2 & 39.3 \\ 
Qwen2-VL-Instruct & 71.9 & 58.5 & 12.6 & 47.7  \\
    \bottomrule
    \end{tabular}
    \caption{Accuracy in Video Question Answering.  D: Descriptive, S: Spatial, A: Adversarial Questions.}
    \label{tab:video_qa}
\end{table}

There have been a lot of works assessing descriptive and spatial understanding of models through video question-answering \citep{yu2019activitynet, xiao2021next, xu2017video, li2024mvbench}, as well as more fine-grained skills like perception and reasoning \citep{patraucean2023perception}, or ego-centric setups \citep{NEURIPS2023_90ce332a}.
Most datasets are compiled from existing corpora \citep{caba2015activitynet, grauman2022ego4d} and crawled from open platforms \citep{thomee2016yfcc100m, shang2019annotating} and do not focus on videos filmed by visually impaired people. 
To address this gap, we curated a new video QA dataset based on videos filmed by BLV users.

\paragraph{Setup} 
We evaluate models on their ability to answer questions given videos recorded by visually impaired people using the object recognition ORBIT \citep{orbit} dataset. We annotate 98 videos and provide 882 question-answer pairs that target three types of questions: 1) descriptive questions regarding the attributes of the objects (colour, shape, number), 2) spatial Understanding about the position of items and their relation to other items, and 3) adversarial questions about items not present in the video \citep{li-etal-2023-evaluating}. Adversarial questions, which cannot be answered based on the information provided in the video, help assess whether models hallucinate responses or can reliably acknowledge uncertainty—a critical safety feature for assistive technologies. More details about the dataset are provided in \cref{appendix:qa}. For evaluation, we follow the LAVE protocol as described above.

\paragraph{Results}
\cref{tab:video_qa} shows the evaluation results for MLLMs that support video inputs. MiniCPM-V-2.6 and LLaVA-Video achieve the highest overall performance, although no model ranks first across all question types.
Notably, we do not observe a performance advantage for models specifically fine-tuned on video data compared to models trained on both images and videos.
Regarding the results per question type, we observe the following patterns. 
While most MLLMs show promising results on descriptive questions, spatial understanding remains challenging even for the best-performing models (VideoChat-Flash and LLaVA-Video at 63-64\%).
Most concerning is the behaviour on adversarial questions, where models consistently provide concrete answers rather than acknowledge uncertainty.
For assistive technologies, this tendency to hallucinate responses instead of expressing an inability to answer could lead to misleading or potentially unsafe guidance \citep{li-etal-2023-evaluating}. In \cref{sec:videoqa-prompt_ablation}, we show that even explicit prompting to express uncertainty as needed yields limited success: while some models improve on adversarial questions, they either achieve only modest gains or overgeneralize uncertain responses to valid questions.

\section{Discussion and Conclusion}
\paragraph{What is missing from existing evaluation frameworks?}
To better understand the use cases and challenges faced by visually impaired individuals, we designed a survey to collect first-hand insights. These findings provide valuable input for designing more effective, user-centred multimodal AI systems and can add evaluation aspects to both targeted and holistic evaluation approaches \citep{liang2023holistic, lee2024vhelm}. Our analysis captured a wide range of challenges that are underexplored or missing from holistic evaluation frameworks, such as 1) technical constraints (latency, real-time settings, internet dependency), 2) multilingual, cultural and contextual understanding, 3) trust and reliability issues amplified by hallucinations, misinterpretations, underspecified responses and failure in safety-critical or ambiguous scenarios.

\paragraph{Can existing models be used as visual assistants?}
We evaluated a range of multimodal models on datasets from visually impaired users, revealing notable limitations. For example, captioning becomes more challenging with culture-specific images, as models struggle to capture cultural nuances and distinctive items. Similarly, in image question answering, models show substantial performance degradation, which aligns with our survey findings. Optical Braille recognition seems to be a new challenge for almost all models, with most failing to perform the task, pointing to gaps in both training data and generalization abilities. In tasks like video object recognition, MLLMs struggle to identify assistive objects, revealing a lack of specificity in recognizing items important to BLV users. For video question answering, models have difficulty answering adversarial questions that refer to items not present in the image, which points to the models' vulnerability in real-world applications where visibility and conditions are not always ideal. 

\paragraph{Beyond classic benchmark evaluation.}
Our findings suggest a pressing need for the development of datasets and models tailored to user needs and preferences. Such datasets should reflect the real-world complexities and unique challenges faced by BLV users across culturally diverse environments, multilingual settings, assistive devices, poor-quality input, and latency constraints. Furthermore, engaging with BLV users in the design and improvement of visual assistants is essential to ensure models address their needs \cite{caselli-etal-2021-guiding, 10.1145/3551624.3555285}. Continuously gathering feedback on usability, accuracy, trust, and preferences can help develop more accessible, contextually aware, and user-centred AI \cite{kirk-etal-2023-past}. 

Traditional evaluation metrics and benchmarks are insufficient in capturing the specific difficulties faced by users in practice \cite{liao2023rethinking, wang-etal-2024-user}. Existing benchmarks focus on general performance and may overlook critical aspects like real-world usability and user satisfaction. To bridge this gap, future research should explore new, reproducible, user-centred methodologies of evaluation that go beyond conventional metrics to better assess models in everyday scenarios \cite{elangovan-etal-2024-considers}. By focusing on the unique challenges of visually impaired individuals and integrating their experiences into the development and evaluation of AI models, we can move towards more effective and inclusive visual assistants.

\section*{Limitations}
While this work offers valuable insights into the potential of MLLMs as visual assistants for the visually impaired, several limitations should be acknowledged. First, our evaluation does not cover tasks related to navigation assistance, which is a crucial aspect of real-world applications for visually impaired individuals. Second, our experimental design focuses primarily on the performance of MLLMs in controlled environments and user-centred tasks and may not fully capture the complexities of dynamic, real-world scenarios. Lastly, our findings underscore the need for further research to address issues related to real-time responsiveness, reasoning tasks, and the inclusion of marginalized languages and cultural contexts.

\section*{Ethics Statement}

This research contributes to the development of AI-driven visual assistants, which have important societal implications, particularly for accessibility and human- computer interaction. We adhere to ethical guidelines. The user survey was carried out with compensation and informed consent, ensuring that participants were fully aware of the purpose of the study and how their data would be used. We took careful measures to protect the privacy and confidentiality of all participants, with no personally identifiable information being disclosed or shared. All datasets used in this work are under CC BY 4.0 license \footnote{\url{https://creativecommons.org/licenses/by/4.0/}}. We acknowledge the potential bias introduced in our survey and evaluation due to the use of datasets and models that may themselves contain inherent biases. Bias could lead to disparities in performance across demographic groups, potentially reinforcing inequalities in AI-assisted technologies. To mitigate these risks, we have taken steps to ensure diversity, such as including participants from different backgrounds in our survey, selecting and adapting use-case specific datasets, evaluating twelve models, and performing qualitative checks of the model outputs. Despite these efforts, inherent biases may persist, and we encourage further scrutiny through external audits and real-world user testing. Future research should focus on developing bias-aware training methods, expanding dataset representativeness, and incorporating user feedback loops to enhance fairness and inclusivity. Additionally, interdisciplinary collaborations with social scientists, ethicists, and affected communities can help refine ethical AI deployment and ensure equitable outcomes for all users. This work emphasizes the need for AI systems that prioritize user trust and safety while acknowledging the potential limitations associated with AI deployment in sensitive contexts.

\section*{Acknowledgements}
The research was supported by the Novo Nordisk Foundation (grant NNF 20SA0066568), and a research grant (VIL53122) from VILLUM FONDEN. Antonia Karamolegkou was supported by the Onassis Foundation - Scholarship ID: F ZP 017-2/2022-2023'. This work was also supported by the Edinburgh International Data Facility (EIDF) and the Data-Driven Innovation Programme at the University of Edinburgh.
The authors acknowledge the use of the HWU high-performance computing facility (DMOG) and associated support services in completing this work. Ruchira Dhar was supported by the Pioneer Centre for AI, DNRF grant number P1.

\bibliography{anthology_reduced,custom}

\begin{thebibliography}{96}
\providecommand{\natexlab}[1]{#1}

\bibitem[{Abdin et~al.(2024)Abdin, Aneja, Awadalla, Awadallah, Awan, Bach, Bahree, Bakhtiari, Bao, Behl et~al.}]{abdin2024phi}
Marah Abdin, Jyoti Aneja, Hany Awadalla, Ahmed Awadallah, Ammar~Ahmad Awan, Nguyen Bach, Amit Bahree, Arash Bakhtiari, Jianmin Bao, Harkirat Behl, et~al. 2024.
\newblock \href {https://arxiv.org/abs/2404.14219} {Phi-3 technical report: A highly capable language model locally on your phone}.
\newblock \emph{arXiv preprint arXiv:2404.14219}.

\bibitem[{Ahsan et~al.(2021)Ahsan, Bhatt, Shah, and Bhalla}]{ahsan-etal-2021-multi}
Hiba Ahsan, Daivat Bhatt, Kaivan Shah, and Nikita Bhalla. 2021.
\newblock \href {https://doi.org/10.18653/v1/2021.naacl-srw.8} {Multi-modal image captioning for the visually impaired}.
\newblock In \emph{Proceedings of the 2021 Conference of the North American Chapter of the Association for Computational Linguistics: Student Research Workshop}, pages 53--60, Online. Association for Computational Linguistics.

\bibitem[{AI@Meta(2024)}]{llama3modelcard}
AI@Meta. 2024.
\newblock \href {https://github.com/meta-llama/llama3/blob/main/MODEL_CARD.md} {Llama 3 model card}.

\bibitem[{Antol et~al.(2015)Antol, Agrawal, Lu, Mitchell, Batra, Zitnick, and Parikh}]{VQA}
Stanislaw Antol, Aishwarya Agrawal, Jiasen Lu, Margaret Mitchell, Dhruv Batra, C.~Lawrence Zitnick, and Devi Parikh. 2015.
\newblock \href {https://openaccess.thecvf.com/content_iccv_2015/html/Antol_VQA_Visual_Question_ICCV_2015_paper.html} {{VQA}: {V}isual {Q}uestion {A}nswering}.
\newblock In \emph{International Conference on Computer Vision (ICCV)}.

\bibitem[{Baker et~al.(2021)Baker, Parekh, Fabre, Addlesee, Kruiper, and Lemon}]{baker-etal-2021-spoon}
Katie Baker, Amit Parekh, Adrien Fabre, Angus Addlesee, Ruben Kruiper, and Oliver Lemon. 2021.
\newblock \href {https://aclanthology.org/2021.reinact-1.5/} {The spoon is in the sink: Assisting visually impaired people in the kitchen}.
\newblock In \emph{Proceedings of the Reasoning and Interaction Conference (ReInAct 2021)}, pages 32--39, Gothenburg, Sweden. Association for Computational Linguistics.

\bibitem[{Beyer et~al.(2024)Beyer, Steiner, Pinto, Kolesnikov, Wang, Salz, Neumann, Alabdulmohsin, Tschannen, Bugliarello et~al.}]{beyer2024paligemma}
Lucas Beyer, Andreas Steiner, Andr{\'e}~Susano Pinto, Alexander Kolesnikov, Xiao Wang, Daniel Salz, Maxim Neumann, Ibrahim Alabdulmohsin, Michael Tschannen, Emanuele Bugliarello, et~al. 2024.
\newblock \href {https://arxiv.org/abs/2407.07726} {Paligemma: A versatile 3b vlm for transfer}.
\newblock \emph{arXiv preprint arXiv:2407.07726}.

\bibitem[{Caselli et~al.(2021)Caselli, Cibin, Conforti, Encinas, and Teli}]{caselli-etal-2021-guiding}
Tommaso Caselli, Roberto Cibin, Costanza Conforti, Enrique Encinas, and Maurizio Teli. 2021.
\newblock \href {https://doi.org/10.18653/v1/2021.nlp4posimpact-1.4} {Guiding principles for participatory design-inspired natural language processing}.
\newblock In \emph{Proceedings of the 1st Workshop on NLP for Positive Impact}, pages 27--35, Online. Association for Computational Linguistics.

\bibitem[{Chen et~al.(2023)Chen, Anjum, and Gurari}]{10377187}
Chongyan Chen, Samreen Anjum, and Danna Gurari. 2023.
\newblock \href {https://doi.org/10.1109/ICCV51070.2023.01405} {Vqa therapy: Exploring answer differences by visually grounding answers}.
\newblock In \emph{2023 IEEE/CVF International Conference on Computer Vision (ICCV)}, pages 15269--15279.

\bibitem[{Chen et~al.(2024)Chen, Sikka, Cogswell, Ji, and Divakaran}]{chen-etal-2024-measuring}
Yangyi Chen, Karan Sikka, Michael Cogswell, Heng Ji, and Ajay Divakaran. 2024.
\newblock \href {https://doi.org/10.18653/v1/2024.naacl-long.11} {Measuring and improving chain-of-thought reasoning in vision-language models}.
\newblock In \emph{Proceedings of the 2024 Conference of the North American Chapter of the Association for Computational Linguistics: Human Language Technologies (Volume 1: Long Papers)}, pages 192--210, Mexico City, Mexico. Association for Computational Linguistics.

\bibitem[{Costa-juss{\`a} et~al.(2022)Costa-juss{\`a}, Cross, {\c{C}}elebi, Elbayad, Heafield, Heffernan, Kalbassi, Lam, Licht, Maillard et~al.}]{costa2022no}
Marta~R Costa-juss{\`a}, James Cross, Onur {\c{C}}elebi, Maha Elbayad, Kenneth Heafield, Kevin Heffernan, Elahe Kalbassi, Janice Lam, Daniel Licht, Jean Maillard, et~al. 2022.
\newblock \href {https://arxiv.org/abs/2207.04672} {No language left behind: Scaling human-centered machine translation}.
\newblock \emph{arXiv preprint arXiv:2207.04672}.

\bibitem[{Damen et~al.(2018)Damen, Doughty, Farinella, Fidler, Furnari, Kazakos, Moltisanti, Munro, Perrett, Price, and Wray}]{Damen_2018_ECCV}
Dima Damen, Hazel Doughty, Giovanni~Maria Farinella, Sanja Fidler, Antonino Furnari, Evangelos Kazakos, Davide Moltisanti, Jonathan Munro, Toby Perrett, Will Price, and Michael Wray. 2018.
\newblock Scaling egocentric vision: The epic-kitchens dataset.
\newblock In \emph{Proceedings of the European Conference on Computer Vision (ECCV)}.

\bibitem[{De~Marsico et~al.(2024)De~Marsico, Giacanelli, Manganaro, Palma, and Santoro}]{de2024vqask}
Maria De~Marsico, Chiara Giacanelli, Clizia~Giorgia Manganaro, Alessio Palma, and Davide Santoro. 2024.
\newblock \href {https://dl.acm.org/doi/10.1145/3656650.3656677} {Vqask: a multimodal android gpt-based application to help blind users visualize pictures}.
\newblock In \emph{Proceedings of the 2024 International Conference on Advanced Visual Interfaces}, pages 1--5.

\bibitem[{Deitke et~al.(2024)Deitke, Clark, Lee, Tripathi, Yang, Park, Salehi, Muennighoff, Lo, Soldaini et~al.}]{deitke2024molmo}
Matt Deitke, Christopher Clark, Sangho Lee, Rohun Tripathi, Yue Yang, Jae~Sung Park, Mohammadreza Salehi, Niklas Muennighoff, Kyle Lo, Luca Soldaini, et~al. 2024.
\newblock \href {https://arxiv.org/abs/2409.17146} {Molmo and pixmo: Open weights and open data for state-of-the-art multimodal models}.
\newblock \emph{arXiv preprint arXiv:2409.17146}.

\bibitem[{Delloul and Larabi(2023)}]{delloul2023real}
Khadidja Delloul and Slimane Larabi. 2023.
\newblock \href {https://arxiv.org/abs/2308.13892} {Towards real time egocentric segment captioning for the blind and visually impaired in rgb-d theatre images}.
\newblock \emph{arXiv preprint}.

\bibitem[{Di~Nuovo et~al.(2024)Di~Nuovo, Sanguinetti, Balestrucci, Anselma, Bernareggi, and Mazzei}]{di-nuovo-etal-2024-educational}
Elisa Di~Nuovo, Manuela Sanguinetti, Pier~Felice Balestrucci, Luca Anselma, Cristian Bernareggi, and Alessandro Mazzei. 2024.
\newblock \href {https://aclanthology.org/2024.lrec-main.489/} {Educational dialogue systems for visually impaired students: Introducing a task-oriented user-agent corpus}.
\newblock In \emph{Proceedings of the 2024 Joint International Conference on Computational Linguistics, Language Resources and Evaluation (LREC-COLING 2024)}, pages 5507--5519, Torino, Italia. ELRA and ICCL.

\bibitem[{Dognin et~al.(2022)Dognin, Melnyk, Mroueh, Padhi, Rigotti, Ross, Schiff, Young, and Belgodere}]{lessons_learned}
Pierre Dognin, Igor Melnyk, Youssef Mroueh, Inkit Padhi, Mattia Rigotti, Jarret Ross, Yair Schiff, Richard~A. Young, and Brian Belgodere. 2022.
\newblock \href {https://doi.org/10.1613/jair.1.13113} {Image captioning as an assistive technology: Lessons learned from vizwiz 2020 challenge}.
\newblock \emph{J. Artif. Int. Res.}, 73.

\bibitem[{Elangovan et~al.(2024)Elangovan, Liu, Xu, Bodapati, and Roth}]{elangovan-etal-2024-considers}
Aparna Elangovan, Ling Liu, Lei Xu, Sravan~Babu Bodapati, and Dan Roth. 2024.
\newblock \href {https://doi.org/10.18653/v1/2024.acl-long.63} {{C}on{S}i{DERS}-the-human evaluation framework: Rethinking human evaluation for generative large language models}.
\newblock In \emph{Proceedings of the 62nd Annual Meeting of the Association for Computational Linguistics (Volume 1: Long Papers)}, pages 1137--1160, Bangkok, Thailand. Association for Computational Linguistics.

\bibitem[{Fabian Caba~Heilbron and Niebles(2015)}]{caba2015activitynet}
Bernard~Ghanem Fabian Caba~Heilbron, Victor~Escorcia and Juan~Carlos Niebles. 2015.
\newblock \href {https://www.cv-foundation.org/openaccess/content_cvpr_2015/papers/Heilbron_ActivityNet_A_Large-Scale_2015_CVPR_paper.pdf} {Activitynet: A large-scale video benchmark for human activity understanding}.
\newblock In \emph{Proceedings of the IEEE Conference on Computer Vision and Pattern Recognition}, pages 961--970.

\bibitem[{Federmann et~al.(2022)Federmann, Kocmi, and Xin}]{federmann-etal-2022-ntrex}
Christian Federmann, Tom Kocmi, and Ying Xin. 2022.
\newblock \href {https://doi.org/10.18653/v1/2022.sumeval-1.4} {{NTREX}-128 {--} news test references for {MT} evaluation of 128 languages}.
\newblock In \emph{Proceedings of the First Workshop on Scaling Up Multilingual Evaluation}, pages 21--24, Online. Association for Computational Linguistics.

\bibitem[{Gao et~al.(2024)Gao, Chang, Ren, Han, and Li}]{gao2024enhanced}
Zhiqiang Gao, Lulu Chang, Bing Ren, Jing Han, and Jie Li. 2024.
\newblock \href {https://www.sciencedirect.com/science/article/abs/pii/S092442472300849X} {Enhanced braille recognition based on piezoresistive and piezoelectric dual-mode tactile sensors}.
\newblock \emph{Sensors and Actuators A: Physical}, 366:115000.

\bibitem[{Geigle et~al.(2024)Geigle, Timofte, and Glava{\v{s}}}]{geigle-etal-2024-african}
Gregor Geigle, Radu Timofte, and Goran Glava{\v{s}}. 2024.
\newblock \href {https://doi.org/10.18653/v1/2024.emnlp-main.154} {{A}frican or {E}uropean swallow? benchmarking large vision-language models for fine-grained object classification}.
\newblock In \emph{Proceedings of the 2024 Conference on Empirical Methods in Natural Language Processing}, pages 2653--2669, Miami, Florida, USA. Association for Computational Linguistics.

\bibitem[{Goyal et~al.(2017{\natexlab{a}})Goyal, Ebrahimi~Kahou, Michalski, Materzynska, Westphal, Kim, Haenel, Fruend, Yianilos, Mueller-Freitag et~al.}]{goyal2017something}
Raghav Goyal, Samira Ebrahimi~Kahou, Vincent Michalski, Joanna Materzynska, Susanne Westphal, Heuna Kim, Valentin Haenel, Ingo Fruend, Peter Yianilos, Moritz Mueller-Freitag, et~al. 2017{\natexlab{a}}.
\newblock \href {https://openaccess.thecvf.com/content_ICCV_2017/papers/Goyal_The_Something_Something_ICCV_2017_paper.pdf} {The" something something" video database for learning and evaluating visual common sense}.
\newblock In \emph{Proceedings of the IEEE international conference on computer vision}, pages 5842--5850.

\bibitem[{Goyal et~al.(2017{\natexlab{b}})Goyal, Khot, Summers-Stay, Batra, and Parikh}]{goyal2017making}
Yash Goyal, Tejas Khot, Douglas Summers-Stay, Dhruv Batra, and Devi Parikh. 2017{\natexlab{b}}.
\newblock \href {https://openaccess.thecvf.com/content_cvpr_2017/html/Goyal_Making_the_v_CVPR_2017_paper.html} {Making the v in vqa matter: Elevating the role of image understanding in visual question answering}.
\newblock In \emph{Proceedings of the IEEE conference on computer vision and pattern recognition}, pages 6904--6913.

\bibitem[{Grauman et~al.(2022)Grauman, Westbury, Byrne, Chavis, Furnari, Girdhar, Hamburger, Jiang, Liu, Liu et~al.}]{grauman2022ego4d}
Kristen Grauman, Andrew Westbury, Eugene Byrne, Zachary Chavis, Antonino Furnari, Rohit Girdhar, Jackson Hamburger, Hao Jiang, Miao Liu, Xingyu Liu, et~al. 2022.
\newblock \href {https://openaccess.thecvf.com/content/CVPR2022/html/Grauman_Ego4D_Around_the_World_in_3000_Hours_of_Egocentric_Video_CVPR_2022_paper.html} {Ego4d: Around the world in 3,000 hours of egocentric video}.
\newblock In \emph{Proceedings of the IEEE/CVF Conference on Computer Vision and Pattern Recognition}, pages 18995--19012.

\bibitem[{Gurari et~al.(2018)Gurari, Li, Stangl, Guo, Lin, Grauman, Luo, and Bigham}]{gurari2018vizwiz}
Danna Gurari, Qing Li, Abigale~J Stangl, Anhong Guo, Chi Lin, Kristen Grauman, Jiebo Luo, and Jeffrey~P Bigham. 2018.
\newblock \href {https://openaccess.thecvf.com/content_cvpr_2018/html/Gurari_VizWiz_Grand_Challenge_CVPR_2018_paper.html} {Vizwiz grand challenge: Answering visual questions from blind people}.
\newblock In \emph{Proceedings of the IEEE conference on computer vision and pattern recognition}, pages 3608--3617.

\bibitem[{Gurari et~al.(2020)Gurari, Zhao, Zhang, and Bhattacharya}]{gurari2020captioning}
Danna Gurari, Yinan Zhao, Meng Zhang, and Nilavra Bhattacharya. 2020.
\newblock \href {https://link.springer.com/chapter/10.1007/978-3-030-58520-4_25} {Captioning images taken by people who are blind}.
\newblock In \emph{Computer Vision--ECCV 2020: 16th European Conference, Glasgow, UK, August 23--28, 2020, Proceedings, Part XVII 16}, pages 417--434. Springer.

\bibitem[{Haddow et~al.(2024)Haddow, Kocmi, Koehn, and Monz}]{wmt-2024-1}
Barry Haddow, Tom Kocmi, Philipp Koehn, and Christof Monz, editors. 2024.
\newblock \href {https://doi.org/10.18653/v1/2024.wmt-1.0} {\emph{Proceedings of the Ninth Conference on Machine Translation}}. Association for Computational Linguistics, Miami, Florida, USA.

\bibitem[{Hessel et~al.(2021)Hessel, Holtzman, Forbes, Le~Bras, and Choi}]{hessel2021clipscore}
Jack Hessel, Ari Holtzman, Maxwell Forbes, Ronan Le~Bras, and Yejin Choi. 2021.
\newblock \href {https://aclanthology.org/2021.emnlp-main.595v2.pdf} {Clipscore: A reference-free evaluation metric for image captioning}.
\newblock In \emph{Proceedings of the 2021 Conference on Empirical Methods in Natural Language Processing}, pages 7514--7528.

\bibitem[{Holiel et~al.(2024)Holiel, Fawzi, and Al-Atabany}]{holiel2024assisting}
Heidi~Ahmed Holiel, Sahar~Ali Fawzi, and Walid Al-Atabany. 2024.
\newblock \href {https://ieeexplore.ieee.org/document/10753262/;jsessionid=3A9C2421CFA53499681EECA562D81B94} {Assisting visually impaired subjects using large language models: A comprehensive evaluation}.
\newblock In \emph{2024 6th Novel Intelligent and Leading Emerging Sciences Conference (NILES)}, pages 561--566. IEEE.

\bibitem[{Hu et~al.(2022)Hu, yelong shen, Wallis, Allen-Zhu, Li, Wang, Wang, and Chen}]{hu2022lora}
Edward~J Hu, yelong shen, Phillip Wallis, Zeyuan Allen-Zhu, Yuanzhi Li, Shean Wang, Lu~Wang, and Weizhu Chen. 2022.
\newblock \href {https://openreview.net/forum?id=nZeVKeeFYf9} {Lo{RA}: Low-rank adaptation of large language models}.
\newblock In \emph{International Conference on Learning Representations}.

\bibitem[{Hu et~al.(2023)Hu, Luan, Chen, Khandelwal, Joshi, Lee, Toutanova, and Chang}]{hu2023open}
Hexiang Hu, Yi~Luan, Yang Chen, Urvashi Khandelwal, Mandar Joshi, Kenton Lee, Kristina Toutanova, and Ming-Wei Chang. 2023.
\newblock \href {https://open-vision-language.github.io/oven/} {Open-domain visual entity recognition: Towards recognizing millions of wikipedia entities}.
\newblock In \emph{Proceedings of the IEEE/CVF International Conference on Computer Vision}, pages 12065--12075.

\bibitem[{Huh et~al.(2024)Huh, Xu, Peng, Chen, Murugu, Gurari, Choi, and Pavel}]{huh2024longform}
Mina Huh, Fangyuan Xu, Yi-Hao Peng, Chongyan Chen, Hansika Murugu, Danna Gurari, Eunsol Choi, and Amy Pavel. 2024.
\newblock \href {https://openreview.net/forum?id=z7FvXbyyrM} {Long-form answers to visual questions from blind and low vision people}.
\newblock In \emph{First Conference on Language Modeling}.

\bibitem[{Islam et~al.(2024)Islam, Kabir, Pearce, Reza, and Billah}]{blv}
Md~Touhidul Islam, Imran Kabir, Elena~Ariel Pearce, Md~Alimoor Reza, and Syed~Masum Billah. 2024.
\newblock \href {https://doi.org/10.1145/3663548.3688538} {Identifying crucial objects in blind and low-vision individuals' navigation}.
\newblock In \emph{Proceedings of the 26th International ACM SIGACCESS Conference on Computers and Accessibility}, ASSETS '24, New York, NY, USA. Association for Computing Machinery.

\bibitem[{Joshi et~al.(2020)Joshi, Santy, Budhiraja, Bali, and Choudhury}]{joshi-etal-2020-state}
Pratik Joshi, Sebastin Santy, Amar Budhiraja, Kalika Bali, and Monojit Choudhury. 2020.
\newblock \href {https://doi.org/10.18653/v1/2020.acl-main.560} {The state and fate of linguistic diversity and inclusion in the {NLP} world}.
\newblock In \emph{Proceedings of the 58th Annual Meeting of the Association for Computational Linguistics}, pages 6282--6293, Online. Association for Computational Linguistics.

\bibitem[{Jung et~al.(2020)Jung, Wada, Crall, Tanaka, Graving, Reinders, Yadav, Banerjee, Vecsei, Kraft, Rui, Borovec, Vallentin, Zhydenko, Pfeiffer, Cook, Fernández, De~Rainville, Weng, Ayala-Acevedo, Meudec, Laporte et~al.}]{imgaug}
Alexander~B. Jung, Kentaro Wada, Jon Crall, Satoshi Tanaka, Jake Graving, Christoph Reinders, Sarthak Yadav, Joy Banerjee, Gábor Vecsei, Adam Kraft, Zheng Rui, Jirka Borovec, Christian Vallentin, Semen Zhydenko, Kilian Pfeiffer, Ben Cook, Ismael Fernández, François-Michel De~Rainville, Chi-Hung Weng, Abner Ayala-Acevedo, Raphael Meudec, Matias Laporte, et~al. 2020.
\newblock {imgaug}.
\newblock \url{https://github.com/aleju/imgaug}.
\newblock Online; accessed 01-Feb-2020.

\bibitem[{Karamolegkou et~al.(2024)Karamolegkou, Rust, Cui, Cao, S{\o}gaard, and Hershcovich}]{karamolegkou-etal-2024-vision}
Antonia Karamolegkou, Phillip Rust, Ruixiang Cui, Yong Cao, Anders S{\o}gaard, and Daniel Hershcovich. 2024.
\newblock \href {https://doi.org/10.18653/v1/2024.hucllm-1.5} {Vision-language models under cultural and inclusive considerations}.
\newblock In \emph{Proceedings of the 1st Human-Centered Large Language Modeling Workshop}, pages 53--66, TBD. ACL.

\bibitem[{Kay et~al.(2017)Kay, Carreira, Simonyan, Zhang, Hillier, Vijayanarasimhan, Viola, Green, Back, Natsev et~al.}]{kay2017kinetics}
Will Kay, Joao Carreira, Karen Simonyan, Brian Zhang, Chloe Hillier, Sudheendra Vijayanarasimhan, Fabio Viola, Tim Green, Trevor Back, Paul Natsev, et~al. 2017.
\newblock \href {https://arxiv.org/abs/1705.06950} {The kinetics human action video dataset}.
\newblock \emph{arXiv preprint arXiv:1705.06950}.

\bibitem[{Kirk et~al.(2023)Kirk, Bean, Vidgen, R{\"o}ttger, and Hale}]{kirk-etal-2023-past}
Hannah~Rose Kirk, Andrew~M. Bean, Bertie Vidgen, Paul R{\"o}ttger, and Scott~A. Hale. 2023.
\newblock \href {https://doi.org/10.18653/v1/2023.emnlp-main.148} {The past, present and better future of feedback learning in large language models for subjective human preferences and values}.
\newblock In \emph{Proceedings of the 2023 Conference on Empirical Methods in Natural Language Processing}, pages 2409--2430, Singapore. Association for Computational Linguistics.

\bibitem[{Kirk et~al.(2024)Kirk, Whitefield, R{\"o}ttger, Bean, Margatina, Mosquera, Ciro, Bartolo, Williams, He, Vidgen, and Hale}]{kirk2024the}
Hannah~Rose Kirk, Alexander Whitefield, Paul R{\"o}ttger, Andrew~Michael Bean, Katerina Margatina, Rafael Mosquera, Juan~Manuel Ciro, Max Bartolo, Adina Williams, He~He, Bertie Vidgen, and Scott~A. Hale. 2024.
\newblock \href {https://openreview.net/forum?id=DFr5hteojx} {The {PRISM} alignment dataset: What participatory, representative and individualised human feedback reveals about the subjective and multicultural alignment of large language models}.
\newblock In \emph{The Thirty-eight Conference on Neural Information Processing Systems Datasets and Benchmarks Track}.

\bibitem[{Lauren{\c{c}}on et~al.(2024)Lauren{\c{c}}on, Marafioti, Sanh, and Tronchon}]{laurenccon2024building}
Hugo Lauren{\c{c}}on, Andr{\'e}s Marafioti, Victor Sanh, and L{\'e}o Tronchon. 2024.
\newblock \href {https://openreview.net/forum?id=iSL0FHZStr} {Building and better understanding vision-language models: insights and future directions}.
\newblock In \emph{Workshop on Responsibly Building the Next Generation of Multimodal Foundational Models}.

\bibitem[{Lee et~al.(2024)Lee, Tu, Wong, Zheng, Zhou, Mai, Roberts, Yasunaga, Yao, Xie, and Liang}]{lee2024vhelm}
Tony Lee, Haoqin Tu, Chi~Heem Wong, Wenhao Zheng, Yiyang Zhou, Yifan Mai, Josselin~Somerville Roberts, Michihiro Yasunaga, Huaxiu Yao, Cihang Xie, and Percy Liang. 2024.
\newblock \href {https://openreview.net/forum?id=TuMnKFKPho} {{VHELM}: A holistic evaluation of vision language models}.
\newblock In \emph{The Thirty-eight Conference on Neural Information Processing Systems Datasets and Benchmarks Track}.

\bibitem[{Li et~al.(2024{\natexlab{a}})Li, Ge, Chen, Ge, Zhang, and Shan}]{li2024seed2plus}
Bohao Li, Yuying Ge, Yi~Chen, Yixiao Ge, Ruimao Zhang, and Ying Shan. 2024{\natexlab{a}}.
\newblock \href {https://arxiv.org/abs/2404.16790} {Seed-bench-2-plus: Benchmarking multimodal large language models with text-rich visual comprehension}.
\newblock \emph{arXiv preprint arXiv:2404.16790}.

\bibitem[{Li et~al.(2024{\natexlab{b}})Li, Ge, Ge, Wang, Wang, Zhang, and Shan}]{Li_2024_CVPR}
Bohao Li, Yuying Ge, Yixiao Ge, Guangzhi Wang, Rui Wang, Ruimao Zhang, and Ying Shan. 2024{\natexlab{b}}.
\newblock \href {https://openaccess.thecvf.com/content/CVPR2024/html/Li_SEED-Bench_Benchmarking_Multimodal_Large_Language_Models_CVPR_2024_paper.html} {Seed-bench: Benchmarking multimodal large language models}.
\newblock In \emph{Proceedings of the IEEE/CVF Conference on Computer Vision and Pattern Recognition (CVPR)}, pages 13299--13308.

\bibitem[{Li et~al.(2024{\natexlab{c}})Li, Wang, He, Li, Wang, Liu, Wang, Xu, Chen, Luo et~al.}]{li2024mvbench}
Kunchang Li, Yali Wang, Yinan He, Yizhuo Li, Yi~Wang, Yi~Liu, Zun Wang, Jilan Xu, Guo Chen, Ping Luo, et~al. 2024{\natexlab{c}}.
\newblock \href {https://dl.acm.org/doi/abs/10.1145/3123266.3123427?casa_token=rSh8y5EgnQ8AAAAA:QXHBTNbm_pLsdnJeNldaeXWjN1Icl5ScWNrKtmpmbt_N7sAXBhrLcmXS5Y1C3PI6ddKma6FadKQ3eCc} {Mvbench: A comprehensive multi-modal video understanding benchmark}.
\newblock In \emph{Proceedings of the IEEE/CVF Conference on Computer Vision and Pattern Recognition}, pages 22195--22206.

\bibitem[{Li et~al.(2020)Li, Liu, Wang, Xu, and Qian}]{li2020optical}
Renqiang Li, Hong Liu, Xiangdong Wang, Jianxing Xu, and Yueliang Qian. 2020.
\newblock \href {https://openaccess.thecvf.com/content_CVPRW_2020/papers/w34/Li_Optical_Braille_Recognition_Based_on_Semantic_Segmentation_Network_With_Auxiliary_CVPRW_2020_paper.pdf} {Optical braille recognition based on semantic segmentation network with auxiliary learning strategy}.
\newblock In \emph{Proceedings of the IEEE/CVF conference on computer vision and pattern recognition workshops}, pages 554--555.

\bibitem[{Li et~al.(2024{\natexlab{d}})Li, Wang, Yu, Zeng, Zhu, Huang, Gao, Li, He, Wang et~al.}]{li2024videochat}
Xinhao Li, Yi~Wang, Jiashuo Yu, Xiangyu Zeng, Yuhan Zhu, Haian Huang, Jianfei Gao, Kunchang Li, Yinan He, Chenting Wang, et~al. 2024{\natexlab{d}}.
\newblock \href {https://arxiv.org/abs/2501.00574} {Videochat-flash: Hierarchical compression for long-context video modeling}.
\newblock \emph{arXiv preprint arXiv:2501.00574}.

\bibitem[{Li et~al.(2023)Li, Du, Zhou, Wang, Zhao, and Wen}]{li-etal-2023-evaluating}
Yifan Li, Yifan Du, Kun Zhou, Jinpeng Wang, Xin Zhao, and Ji-Rong Wen. 2023.
\newblock \href {https://doi.org/10.18653/v1/2023.emnlp-main.20} {Evaluating object hallucination in large vision-language models}.
\newblock In \emph{Proceedings of the 2023 Conference on Empirical Methods in Natural Language Processing}, pages 292--305, Singapore. Association for Computational Linguistics.

\bibitem[{Liang et~al.(2023)Liang, Bommasani, Lee, Tsipras, Soylu, Yasunaga, Zhang, Narayanan, Wu, Kumar, Newman, Yuan, Yan, Zhang, Cosgrove, Manning, Re, Acosta-Navas, Hudson, Zelikman, Durmus, Ladhak, Rong, Ren, Yao, WANG, Santhanam, Orr, Zheng, Yuksekgonul, Suzgun, Kim, Guha, Chatterji, Khattab, Henderson, Huang, Chi, Xie, Santurkar, Ganguli, Hashimoto, Icard, Zhang, Chaudhary, Wang, Li, Mai, Zhang, and Koreeda}]{liang2023holistic}
Percy Liang, Rishi Bommasani, Tony Lee, Dimitris Tsipras, Dilara Soylu, Michihiro Yasunaga, Yian Zhang, Deepak Narayanan, Yuhuai Wu, Ananya Kumar, Benjamin Newman, Binhang Yuan, Bobby Yan, Ce~Zhang, Christian~Alexander Cosgrove, Christopher~D Manning, Christopher Re, Diana Acosta-Navas, Drew~Arad Hudson, Eric Zelikman, Esin Durmus, Faisal Ladhak, Frieda Rong, Hongyu Ren, Huaxiu Yao, Jue WANG, Keshav Santhanam, Laurel Orr, Lucia Zheng, Mert Yuksekgonul, Mirac Suzgun, Nathan Kim, Neel Guha, Niladri~S. Chatterji, Omar Khattab, Peter Henderson, Qian Huang, Ryan~Andrew Chi, Sang~Michael Xie, Shibani Santurkar, Surya Ganguli, Tatsunori Hashimoto, Thomas Icard, Tianyi Zhang, Vishrav Chaudhary, William Wang, Xuechen Li, Yifan Mai, Yuhui Zhang, and Yuta Koreeda. 2023.
\newblock \href {https://openreview.net/forum?id=iO4LZibEqW} {Holistic evaluation of language models}.
\newblock \emph{Transactions on Machine Learning Research}.
\newblock Featured Certification, Expert Certification.

\bibitem[{Liao and Xiao(2023)}]{liao2023rethinking}
Q~Vera Liao and Ziang Xiao. 2023.
\newblock \href {https://arxiv.org/abs/2306.03100} {Rethinking model evaluation as narrowing the socio-technical gap}.
\newblock \emph{arXiv preprint arXiv:2306.03100}.

\bibitem[{Liao et~al.(2024)Liao, Antoniak, Cheong, Cheng, Lee, Lo, Chang, and Zhang}]{researchtools}
Zhehui Liao, Maria Antoniak, Inyoung Cheong, Evie Yu-Yen Cheng, Ai-Heng Lee, Kyle Lo, Joseph~Chee Chang, and Amy~X. Zhang. 2024.
\newblock \href {https://arxiv.org/abs/2411.05025} {Llms as research tools: A large scale survey of researchers' usage and perceptions}.
\newblock \emph{Preprint}, arXiv:2411.05025.

\bibitem[{Liu et~al.(2024{\natexlab{a}})Liu, Li, Li, and Lee}]{liu2024improved}
Haotian Liu, Chunyuan Li, Yuheng Li, and Yong~Jae Lee. 2024{\natexlab{a}}.
\newblock \href {https://openaccess.thecvf.com/content/CVPR2024/html/Liu_Improved_Baselines_with_Visual_Instruction_Tuning_CVPR_2024_paper.html} {Improved baselines with visual instruction tuning}.
\newblock In \emph{Proceedings of the IEEE/CVF Conference on Computer Vision and Pattern Recognition}, pages 26296--26306.

\bibitem[{Liu et~al.(2024{\natexlab{b}})Liu, Yang, Zhong, Tholeti, Ding, Zhang, and Gilpin}]{liu2024right}
Li~Liu, Diji Yang, Sijia Zhong, Kalyana Suma~Sree Tholeti, Lei Ding, Yi~Zhang, and Leilani~H. Gilpin. 2024{\natexlab{b}}.
\newblock \href {https://openreview.net/forum?id=7ANmKBfP88} {Right this way: Can {VLM}s guide us to see more to answer questions?}
\newblock In \emph{The Thirty-eighth Annual Conference on Neural Information Processing Systems}.

\bibitem[{Liu et~al.(2024{\natexlab{c}})Liu, Duan, Zhang, Li, Zhang, Zhao, Yuan, Wang, He, Liu et~al.}]{liu2024mmbench}
Yuan Liu, Haodong Duan, Yuanhan Zhang, Bo~Li, Songyang Zhang, Wangbo Zhao, Yike Yuan, Jiaqi Wang, Conghui He, Ziwei Liu, et~al. 2024{\natexlab{c}}.
\newblock \href {https://link.springer.com/chapter/10.1007/978-3-031-72658-3_13} {Mmbench: Is your multi-modal model an all-around player?}
\newblock In \emph{European conference on computer vision}, pages 216--233. Springer.

\bibitem[{Liu et~al.(2024{\natexlab{d}})Liu, Li, Huang, Yang, Yu, Li, Yin, Liu, Jin, and Bai}]{Liu_2024}
Yuliang Liu, Zhang Li, Mingxin Huang, Biao Yang, Wenwen Yu, Chunyuan Li, Xu-Cheng Yin, Cheng-Lin Liu, Lianwen Jin, and Xiang Bai. 2024{\natexlab{d}}.
\newblock \href {https://doi.org/10.1007/s11432-024-4235-6} {Ocrbench: on the hidden mystery of ocr in large multimodal models}.
\newblock \emph{Science China Information Sciences}, 67(12).

\bibitem[{Lomonaco and Maltoni(2017)}]{core}
Vincenzo Lomonaco and Davide Maltoni. 2017.
\newblock \href {https://proceedings.mlr.press/v78/lomonaco17a.html} {Core50: a new dataset and benchmark for continuous object recognition}.
\newblock In \emph{Proceedings of the 1st Annual Conference on Robot Learning}, volume~78 of \emph{Proceedings of Machine Learning Research}, pages 17--26. PMLR.

\bibitem[{Lu et~al.(2024)Lu, Bansal, Xia, Liu, Li, Hajishirzi, Cheng, Chang, Galley, and Gao}]{lu2024mathvista}
Pan Lu, Hritik Bansal, Tony Xia, Jiacheng Liu, Chunyuan Li, Hannaneh Hajishirzi, Hao Cheng, Kai-Wei Chang, Michel Galley, and Jianfeng Gao. 2024.
\newblock \href {https://openreview.net/forum?id=KUNzEQMWU7} {Mathvista: Evaluating mathematical reasoning of foundation models in visual contexts}.
\newblock In \emph{The Twelfth International Conference on Learning Representations}.

\bibitem[{Ma{\~n}as et~al.(2024)Ma{\~n}as, Krojer, and Agrawal}]{manas2024improving}
Oscar Ma{\~n}as, Benno Krojer, and Aishwarya Agrawal. 2024.
\newblock \href {https://arxiv.org/abs/2310.02567} {Improving automatic vqa evaluation using large language models}.
\newblock In \emph{Proceedings of the AAAI Conference on Artificial Intelligence}, 5, pages 4171--4179.

\bibitem[{Mangalam et~al.(2023)Mangalam, Akshulakov, and Malik}]{NEURIPS2023_90ce332a}
Karttikeya Mangalam, Raiymbek Akshulakov, and Jitendra Malik. 2023.
\newblock \href {https://proceedings.neurips.cc/paper_files/paper/2023/hash/90ce332aff156b910b002ce4e6880dec-Abstract-Datasets_and_Benchmarks.html} {Egoschema: A diagnostic benchmark for very long-form video language understanding}.
\newblock In \emph{Advances in Neural Information Processing Systems}, volume~36, pages 46212--46244. Curran Associates, Inc.

\bibitem[{Massiceti et~al.(2021)Massiceti, Zintgraf, Bronskill, Theodorou, Harris, Cutrell, Morrison, Hofmann, and Stumpf}]{orbit}
Daniela Massiceti, Luisa Zintgraf, John Bronskill, Lida Theodorou, Matthew~Tobias Harris, Edward Cutrell, Cecily Morrison, Katja Hofmann, and Simone Stumpf. 2021.
\newblock \href {https://openaccess.thecvf.com/content/ICCV2021/papers/Massiceti_ORBIT_A_Real-World_Few-Shot_Dataset_for_Teachable_Object_Recognition_ICCV_2021_paper.pdf} {{ORBIT: A Real-World Few-Shot Dataset for Teachable Object Recognition}}.
\newblock In \emph{Proceedings of the IEEE/CVF International Conference on Computer Vision (ICCV)}.

\bibitem[{{Meta}(2024)}]{llama3.2}
{Meta}. 2024.
\newblock \href {https://ai.meta.com/blog/llama-3-2-connect-2024-vision-edge-mobile-devices/} {Llama 3.2: Revolutionizing edge ai and vision with open, customizable models}.

\bibitem[{Mogrovejo et~al.(2024)Mogrovejo, Lyu, Wibowo, G{\'o}ngora, Mandal, Purkayastha, Ortiz-Barajas, Cueva, Baek, Jeong, Hamed, Yong, Lim, Silva, Dunstan, Jouitteau, MEUR, Nwatu, Batnasan, Otgonbold, Gochoo, Ivetta, Benotti, Alemany, Maina, Geng, Torrent, Belcavello, Viridiano, Cruz, Velasco, Ignat, Burzo, Whitehouse, Abzaliev, Clifford, Caulfield, Lynn, Salamea-Palacios, Araujo, Kementchedjhieva, Mihaylov, Azime, Ademtew, Balcha, Etori, Adelani, Mihalcea, Tonja, Cabrera, Vallejo, Lovenia, Zhang, Estecha-Garitagoitia, Rodr{\'\i}guez-Cantelar, Ehsan, Chevi, Adilazuarda, Diandaru, Cahyawijaya, Koto, Kuribayashi, Song, Khandavally, Jayakumar, Dabre, Imam, Nagasinghe, Dragonetti, D'Haro, NIYOMUGISHA, Gala, Chitale, Farooqui, Solorio, and Aji}]{mogrovejo2024cvqa}
David Orlando~Romero Mogrovejo, Chenyang Lyu, Haryo~Akbarianto Wibowo, Santiago G{\'o}ngora, Aishik Mandal, Sukannya Purkayastha, Jesus-German Ortiz-Barajas, Emilio~Villa Cueva, Jinheon Baek, Soyeong Jeong, Injy Hamed, Zheng~Xin Yong, Zheng~Wei Lim, Paula~M{\'o}nica Silva, Jocelyn Dunstan, M{\'e}lanie Jouitteau, David~LE MEUR, Joan Nwatu, Ganzorig Batnasan, Munkh-Erdene Otgonbold, Munkhjargal Gochoo, Guido Ivetta, Luciana Benotti, Laura~Alonso Alemany, Hern{\'a}n Maina, Jiahui Geng, Tiago~Timponi Torrent, Frederico Belcavello, Marcelo Viridiano, Jan Christian~Blaise Cruz, Dan~John Velasco, Oana Ignat, Zara Burzo, Chenxi Whitehouse, Artem Abzaliev, Teresa Clifford, Gr{\'a}inne Caulfield, Teresa Lynn, Christian Salamea-Palacios, Vladimir Araujo, Yova Kementchedjhieva, Mihail~Minkov Mihaylov, Israel~Abebe Azime, Henok~Biadglign Ademtew, Bontu~Fufa Balcha, Naome~A Etori, David~Ifeoluwa Adelani, Rada Mihalcea, Atnafu~Lambebo Tonja, Maria Camila~Buitrago Cabrera, Gisela Vallejo, Holy Lovenia, Ruochen Zhang, Marcos
  Estecha-Garitagoitia, Mario Rodr{\'\i}guez-Cantelar, Toqeer Ehsan, Rendi Chevi, Muhammad~Farid Adilazuarda, Ryandito Diandaru, Samuel Cahyawijaya, Fajri Koto, Tatsuki Kuribayashi, Haiyue Song, Aditya Nanda~Kishore Khandavally, Thanmay Jayakumar, Raj Dabre, Mohamed Fazli~Mohamed Imam, Kumaranage Ravindu~Yasas Nagasinghe, Alina Dragonetti, Luis~Fernando D'Haro, Olivier NIYOMUGISHA, Jay Gala, Pranjal~A Chitale, Fauzan Farooqui, Thamar Solorio, and Alham~Fikri Aji. 2024.
\newblock \href {https://openreview.net/forum?id=E18kRXTGmV} {{CVQA}: Culturally-diverse multilingual visual question answering benchmark}.
\newblock In \emph{The Thirty-eight Conference on Neural Information Processing Systems Datasets and Benchmarks Track}.

\bibitem[{Nayak et~al.(2024)Nayak, Jain, Awal, Reddy, Steenkiste, Hendricks, Stanczak, and Agrawal}]{nayak-etal-2024-benchmarking}
Shravan Nayak, Kanishk Jain, Rabiul Awal, Siva Reddy, Sjoerd~Van Steenkiste, Lisa~Anne Hendricks, Karolina Stanczak, and Aishwarya Agrawal. 2024.
\newblock \href {https://doi.org/10.18653/v1/2024.emnlp-main.329} {Benchmarking vision language models for cultural understanding}.
\newblock In \emph{Proceedings of the 2024 Conference on Empirical Methods in Natural Language Processing}, pages 5769--5790, Miami, Florida, USA. Association for Computational Linguistics.

\bibitem[{{NLLB Team} et~al.(2024){NLLB Team}, Costa-juss{\`a}, Cross, {\c{C}}elebi, Elbayad, Heafield, Heffernan, Kalbassi, Lam, Licht, Maillard, Sun, Wang, Wenzek, Youngblood, Akula, Barrault, Gonzalez, Hansanti, Hoffman, Jarrett, Sadagopan, Rowe, Spruit, Tran, Andrews, Ayan, Bhosale, Edunov, Fan, Gao, Goswami, Guzm{\'a}n, Koehn, Mourachko, Ropers, Saleem, Schwenk, and Wang}]{nllb-24}
{NLLB Team}, Marta~R. Costa-juss{\`a}, James Cross, Onur {\c{C}}elebi, Maha Elbayad, Kenneth Heafield, Kevin Heffernan, Elahe Kalbassi, Janice Lam, Daniel Licht, Jean Maillard, Anna Sun, Skyler Wang, Guillaume Wenzek, Al~Youngblood, Bapi Akula, Loic Barrault, Gabriel~Mejia Gonzalez, Prangthip Hansanti, John Hoffman, Semarley Jarrett, Kaushik~Ram Sadagopan, Dirk Rowe, Shannon Spruit, Chau Tran, Pierre Andrews, Necip~Fazil Ayan, Shruti Bhosale, Sergey Edunov, Angela Fan, Cynthia Gao, Vedanuj Goswami, Francisco Guzm{\'a}n, Philipp Koehn, Alexandre Mourachko, Christophe Ropers, Safiyyah Saleem, Holger Schwenk, and Jeff Wang. 2024.
\newblock \href {https://doi.org/10.1038/s41586-024-07335-x} {Scaling neural machine translation to 200 languages}.
\newblock \emph{Nature}, 630(8018):841--846.

\bibitem[{Papineni et~al.(2002)Papineni, Roukos, Ward, and Zhu}]{papineni2002bleu}
Kishore Papineni, Salim Roukos, Todd Ward, and Wei-Jing Zhu. 2002.
\newblock \href {https://aclanthology.org/P02-1040.pdf} {Bleu: a method for automatic evaluation of machine translation}.
\newblock In \emph{Proceedings of the 40th annual meeting of the Association for Computational Linguistics}, pages 311--318.

\bibitem[{Parashar et~al.(2024)Parashar, Lin, Liu, Dong, Li, Ramanan, Caverlee, and Kong}]{parashar2024neglected}
Shubham Parashar, Zhiqiu Lin, Tian Liu, Xiangjue Dong, Yanan Li, Deva Ramanan, James Caverlee, and Shu Kong. 2024.
\newblock \href {https://openaccess.thecvf.com/content/CVPR2024/html/Parashar_The_Neglected_Tails_in_Vision-Language_Models_CVPR_2024_paper.html} {The neglected tails in vision-language models}.
\newblock In \emph{Proceedings of the IEEE/CVF Conference on Computer Vision and Pattern Recognition}, pages 12988--12997.

\bibitem[{Patraucean et~al.(2023)Patraucean, Smaira, Gupta, Continente, Markeeva, Banarse, Koppula, Heyward, Malinowski, Yang, Doersch, Matejovicova, Sulsky, Miech, Fr{\'e}chette, Klimczak, Koster, Zhang, Winkler, Aytar, Osindero, Damen, Zisserman, and Carreira}]{patraucean2023perception}
Viorica Patraucean, Lucas Smaira, Ankush Gupta, Adria~Recasens Continente, Larisa Markeeva, Dylan~Sunil Banarse, Skanda Koppula, Joseph Heyward, Mateusz Malinowski, Yi~Yang, Carl Doersch, Tatiana Matejovicova, Yury Sulsky, Antoine Miech, Alexandre Fr{\'e}chette, Hanna Klimczak, Raphael Koster, Junlin Zhang, Stephanie Winkler, Yusuf Aytar, Simon Osindero, Dima Damen, Andrew Zisserman, and Joao Carreira. 2023.
\newblock \href {https://openreview.net/forum?id=HYEGXFnPoq} {Perception test: A diagnostic benchmark for multimodal video models}.
\newblock In \emph{Thirty-seventh Conference on Neural Information Processing Systems Datasets and Benchmarks Track}.

\bibitem[{Pfeiffer et~al.(2022)Pfeiffer, Geigle, Kamath, Steitz, Roth, Vuli{\'c}, and Gurevych}]{pfeiffer-etal-2022-xgqa}
Jonas Pfeiffer, Gregor Geigle, Aishwarya Kamath, Jan-Martin~O. Steitz, Stefan Roth, Ivan Vuli{\'c}, and Iryna Gurevych. 2022.
\newblock \href {https://doi.org/10.18653/v1/2022.findings-acl.196} {x{GQA}: Cross-lingual visual question answering}.
\newblock In \emph{Findings of the Association for Computational Linguistics: ACL 2022}, pages 2497--2511, Dublin, Ireland. Association for Computational Linguistics.

\bibitem[{Popovi{\'c}(2017)}]{popovic-2017-chrf}
Maja Popovi{\'c}. 2017.
\newblock \href {https://doi.org/10.18653/v1/W17-4770} {chr{F}++: words helping character n-grams}.
\newblock In \emph{Proceedings of the Second Conference on Machine Translation}, pages 612--618, Copenhagen, Denmark. Association for Computational Linguistics.

\bibitem[{Rajpurkar et~al.(2018)Rajpurkar, Jia, and Liang}]{rajpurkar-etal-2018-know}
Pranav Rajpurkar, Robin Jia, and Percy Liang. 2018.
\newblock \href {https://doi.org/10.18653/v1/P18-2124} {Know what you don`t know: Unanswerable questions for {SQ}u{AD}}.
\newblock In \emph{Proceedings of the 56th Annual Meeting of the Association for Computational Linguistics (Volume 2: Short Papers)}, pages 784--789, Melbourne, Australia. Association for Computational Linguistics.

\bibitem[{Reynolds et~al.(2024)Reynolds, Nagesh, and Gurari}]{reynolds2024salient}
Jarek Reynolds, Chandra~Kanth Nagesh, and Danna Gurari. 2024.
\newblock \href {https://openaccess.thecvf.com/content/WACV2024/html/Reynolds_Salient_Object_Detection_for_Images_Taken_by_People_With_Vision_WACV_2024_paper.html} {Salient object detection for images taken by people with vision impairments}.
\newblock In \emph{Proceedings of the IEEE/CVF Winter Conference on Applications of Computer Vision}, pages 8522--8531.

\bibitem[{Russakovsky et~al.(2015)Russakovsky, Deng, Su, Krause, Satheesh, Ma, Huang, Karpathy, Khosla, Bernstein, Berg, and Fei-Fei}]{ILSVRC15}
Olga Russakovsky, Jia Deng, Hao Su, Jonathan Krause, Sanjeev Satheesh, Sean Ma, Zhiheng Huang, Andrej Karpathy, Aditya Khosla, Michael Bernstein, Alexander~C. Berg, and Li~Fei-Fei. 2015.
\newblock \href {https://doi.org/10.1007/s11263-015-0816-y} {{ImageNet Large Scale Visual Recognition Challenge}}.
\newblock \emph{International Journal of Computer Vision (IJCV)}, 115(3):211--252.

\bibitem[{Schwenk et~al.(2022)Schwenk, Khandelwal, Clark, Marino, and Mottaghi}]{schwenk2022okvqa}
Dustin Schwenk, Apoorv Khandelwal, Christopher Clark, Kenneth Marino, and Roozbeh Mottaghi. 2022.
\newblock \href {https://link.springer.com/chapter/10.1007/978-3-031-20074-8_9} {A-okvqa: A benchmark for visual question answering using world knowledge}.
\newblock In \emph{European conference on computer vision}, pages 146--162. Springer.

\bibitem[{Shang et~al.(2019)Shang, Di, Xiao, Cao, Yang, and Chua}]{shang2019annotating}
Xindi Shang, Donglin Di, Junbin Xiao, Yu~Cao, Xun Yang, and Tat-Seng Chua. 2019.
\newblock \href {https://dl.acm.org/doi/10.1145/3323873.3325056} {Annotating objects and relations in user-generated videos}.
\newblock In \emph{Proceedings of the 2019 on International Conference on Multimedia Retrieval}, pages 279--287. ACM.

\bibitem[{Sloane et~al.(2022)Sloane, Moss, Awomolo, and Forlano}]{10.1145/3551624.3555285}
Mona Sloane, Emanuel Moss, Olaitan Awomolo, and Laura Forlano. 2022.
\newblock \href {https://doi.org/10.1145/3551624.3555285} {Participation is not a design fix for machine learning}.
\newblock In \emph{Proceedings of the 2nd ACM Conference on Equity and Access in Algorithms, Mechanisms, and Optimization}, EAAMO '22, New York, NY, USA. Association for Computing Machinery.

\bibitem[{Smelyakov et~al.(2018)Smelyakov, Chupryna, Yeremenko, Sakhon, and Polezhai}]{smelyakov2018Braille}
Kirill Smelyakov, Anastasiya Chupryna, Dmytro Yeremenko, Anton Sakhon, and Vitalii Polezhai. 2018.
\newblock \href {https://ieeexplore.ieee.org/document/8478615} {Braille character recognition based on neural networks}.
\newblock In \emph{2018 IEEE Second International Conference on Data Stream Mining \& Processing (DSMP)}, pages 509--513. IEEE.

\bibitem[{Sun et~al.(2021)Sun, Yao, Wei, Zhang, Shen, Wu, Zhang, and Shen}]{webly}
Zeren Sun, Yazhou Yao, Xiu-Shen Wei, Yongshun Zhang, Fumin Shen, Jianxin Wu, Jian Zhang, and Heng~Tao Shen. 2021.
\newblock \href {https://github.com/NUST-Machine-Intelligence-Laboratory/weblyFG-dataset} {Webly supervised fine-grained recognition: Benchmark datasets and an approach}.
\newblock In \emph{IEEE International Conference on Computer Vision (ICCV)}.

\bibitem[{Thapliyal et~al.(2022)Thapliyal, Pont-Tuset, Chen, and Soricut}]{ThapliyalCrossmodal2022}
Ashish Thapliyal, Jordi Pont-Tuset, Xi~Chen, and Radu Soricut. 2022.
\newblock \href {https://aclanthology.org/2022.emnlp-main.45/} {{Crossmodal-3600: A Massively Multilingual Multimodal Evaluation Dataset}}.
\newblock In \emph{EMNLP}.

\bibitem[{Thomee et~al.(2016)Thomee, Shamma, Friedland, Elizalde, Ni, Poland, Borth, and Li}]{thomee2016yfcc100m}
Bart Thomee, David~A Shamma, Gerald Friedland, Benjamin Elizalde, Karl Ni, Douglas Poland, Damian Borth, and Li-Jia Li. 2016.
\newblock \href {https://dl.acm.org/doi/abs/10.1145/2812802} {Yfcc100m: The new data in multimedia research}.
\newblock \emph{Communications of the ACM}, 59(2):64--73.

\bibitem[{Tong et~al.(2024)Tong, Liu, Zhai, Ma, LeCun, and Xie}]{eyes_shut}
Shengbang Tong, Zhuang Liu, Yuexiang Zhai, Yi~Ma, Yann LeCun, and Saining Xie. 2024.
\newblock \href {https://doi.org/10.1109/CVPR52733.2024.00914} {Eyes wide shut? exploring the visual shortcomings of multimodal llms}.
\newblock In \emph{2024 IEEE/CVF Conference on Computer Vision and Pattern Recognition (CVPR)}, pages 9568--9578.

\bibitem[{Tseng et~al.(2022)Tseng, Bell, and Gurari}]{tseng2022vizwiz}
Yu-Yun Tseng, Alexander Bell, and Danna Gurari. 2022.
\newblock \href {https://link.springer.com/chapter/10.1007/978-3-031-20074-8_33} {Vizwiz-fewshot: Locating objects in images taken by people with visual impairments}.
\newblock In \emph{European Conference on Computer Vision}, pages 575--591. Springer.

\bibitem[{Tseng et~al.(2024)Tseng, Sharma, Zhang, Stangl, Findlater, Wang, and Gurari}]{tseng2024biv}
Yu-Yun Tseng, Tanusree Sharma, Lotus Zhang, Abigale Stangl, Leah Findlater, Yang Wang, and Danna Gurari. 2024.
\newblock \href {https://arxiv.org/abs/2407.18243} {Biv-priv-seg: Locating private content in images taken by people with visual impairments}.
\newblock \emph{arXiv preprint arXiv:2407.18243}.

\bibitem[{Wang et~al.(2024{\natexlab{a}})Wang, Mo, Ma, Sun, Zhang, and Nie}]{wang-etal-2024-user}
Jiayin Wang, Fengran Mo, Weizhi Ma, Peijie Sun, Min Zhang, and Jian-Yun Nie. 2024{\natexlab{a}}.
\newblock \href {https://doi.org/10.18653/v1/2024.emnlp-main.210} {A user-centric multi-intent benchmark for evaluating large language models}.
\newblock In \emph{Proceedings of the 2024 Conference on Empirical Methods in Natural Language Processing}, pages 3588--3612, Miami, Florida, USA. Association for Computational Linguistics.

\bibitem[{Wang et~al.(2024{\natexlab{b}})Wang, Pan, Shi, Lu, Ren, Zhou, Zhan, and Li}]{wang2024measuring}
Ke~Wang, Junting Pan, Weikang Shi, Zimu Lu, Houxing Ren, Aojun Zhou, Mingjie Zhan, and Hongsheng Li. 2024{\natexlab{b}}.
\newblock \href {https://openreview.net/forum?id=QWTCcxMpPA} {Measuring multimodal mathematical reasoning with {MATH}-vision dataset}.
\newblock In \emph{The Thirty-eight Conference on Neural Information Processing Systems Datasets and Benchmarks Track}.

\bibitem[{Wang et~al.(2024{\natexlab{c}})Wang, Bai, Tan, Wang, Fan, Bai, Chen, Liu, Wang, Ge et~al.}]{wang2024qwen2}
Peng Wang, Shuai Bai, Sinan Tan, Shijie Wang, Zhihao Fan, Jinze Bai, Keqin Chen, Xuejing Liu, Jialin Wang, Wenbin Ge, et~al. 2024{\natexlab{c}}.
\newblock \href {https://arxiv.org/abs/2409.12191} {Qwen2-vl: Enhancing vision-language model's perception of the world at any resolution}.
\newblock \emph{arXiv preprint arXiv:2409.12191}.

\bibitem[{Wang et~al.(2024{\natexlab{d}})Wang, Chen, Wang, Cao, Liu, Gao, Zhu, Zhu, Lu, Qiao et~al.}]{wang2024enhancing}
Weiyun Wang, Zhe Chen, Wenhai Wang, Yue Cao, Yangzhou Liu, Zhangwei Gao, Jinguo Zhu, Xizhou Zhu, Lewei Lu, Yu~Qiao, et~al. 2024{\natexlab{d}}.
\newblock \href {https://arxiv.org/abs/2411.10442} {Enhancing the reasoning ability of multimodal large language models via mixed preference optimization}.
\newblock \emph{arXiv preprint arXiv:2411.10442}.

\bibitem[{Xiao et~al.(2021)Xiao, Shang, Yao, and Chua}]{xiao2021next}
Junbin Xiao, Xindi Shang, Angela Yao, and Tat-Seng Chua. 2021.
\newblock \href {https://arxiv.org/pdf/2105.08276} {Next-qa: Next phase of question-answering to explaining temporal actions}.
\newblock In \emph{Proceedings of the IEEE/CVF conference on computer vision and pattern recognition}, pages 9777--9786.

\bibitem[{Xu et~al.(2017)Xu, Zhao, Xiao, Wu, Zhang, He, and Zhuang}]{xu2017video}
Dejing Xu, Zhou Zhao, Jun Xiao, Fei Wu, Hanwang Zhang, Xiangnan He, and Yueting Zhuang. 2017.
\newblock \href {https://dl.acm.org/doi/abs/10.1145/3123266.3123427?casa_token=rSh8y5EgnQ8AAAAA:QXHBTNbm_pLsdnJeNldaeXWjN1Icl5ScWNrKtmpmbt_N7sAXBhrLcmXS5Y1C3PI6ddKma6FadKQ3eCc} {Video question answering via gradually refined attention over appearance and motion}.
\newblock In \emph{Proceedings of the 25th ACM international conference on Multimedia}, pages 1645--1653.

\bibitem[{Yang et~al.(2024)Yang, He, Liu, and Yan}]{yang2024viassist}
Bufang Yang, Lixing He, Kaiwei Liu, and Zhenyu Yan. 2024.
\newblock \href {https://arxiv.org/abs/2404.02508} {Viassist: Adapting multi-modal large language models for users with visual impairments}.
\newblock \emph{arXiv preprint arXiv:2404.02508}.

\bibitem[{Yao et~al.(2024)Yao, Yu, Zhang, Wang, Cui, Zhu, Cai, Li, Zhao, He et~al.}]{yao2024minicpm}
Yuan Yao, Tianyu Yu, Ao~Zhang, Chongyi Wang, Junbo Cui, Hongji Zhu, Tianchi Cai, Haoyu Li, Weilin Zhao, Zhihui He, et~al. 2024.
\newblock \href {https://arxiv.org/abs/2408.01800} {Minicpm-v: A gpt-4v level mllm on your phone}.
\newblock \emph{arXiv preprint arXiv:2408.01800}.

\bibitem[{Yin et~al.(2023)Yin, Fu, Zhao, Li, Sun, Xu, and Chen}]{yin2023survey}
Shukang Yin, Chaoyou Fu, Sirui Zhao, Ke~Li, Xing Sun, Tong Xu, and Enhong Chen. 2023.
\newblock \href {https://arxiv.org/abs/2306.13549} {A survey on multimodal large language models}.
\newblock \emph{arXiv preprint arXiv:2306.13549}.

\bibitem[{Yu et~al.(2023)Yu, Nikandrou, Jin, and Rieser}]{yu2023quality}
Lu~Yu, Malvina Nikandrou, Jiali Jin, and Verena Rieser. 2023.
\newblock \href {https://dl.acm.org/doi/abs/10.24963/ijcai.2023/697} {Quality-agnostic image captioning to safely assist people with vision impairment}.
\newblock In \emph{Proceedings of the Thirty-Second International Joint Conference on Artificial Intelligence}, pages 6281--6289.

\bibitem[{Yu et~al.(2019)Yu, Xu, Yu, Yu, Zhao, Zhuang, and Tao}]{yu2019activitynet}
Zhou Yu, Dejing Xu, Jun Yu, Ting Yu, Zhou Zhao, Yueting Zhuang, and Dacheng Tao. 2019.
\newblock \href {https://ojs.aaai.org/index.php/AAAI/article/view/4946} {Activitynet-qa: A dataset for understanding complex web videos via question answering}.
\newblock In \emph{Proceedings of the AAAI Conference on Artificial Intelligence}, 01, pages 9127--9134.

\bibitem[{Yuan et~al.(2025)Yuan, Zhang, Zhang, Zhou, and Zhang}]{yuan2025walkvlmaidvisuallyimpairedpeople}
Zhiqiang Yuan, Ting Zhang, Jiapei Zhang, Jie Zhou, and Jinchao Zhang. 2025.
\newblock \href {https://arxiv.org/abs/2412.20903} {Walkvlm:aid visually impaired people walking by vision language model}.
\newblock \emph{Preprint}, arXiv:2412.20903.

\bibitem[{Yue et~al.(2024)Yue, Song, Asai, Kim, Nyandwi, Khanuja, Kantharuban, Sutawika, Ramamoorthy, and Neubig}]{yue2024pangea}
Xiang Yue, Yueqi Song, Akari Asai, Seungone Kim, Jean de~Dieu Nyandwi, Simran Khanuja, Anjali Kantharuban, Lintang Sutawika, Sathyanarayanan Ramamoorthy, and Graham Neubig. 2024.
\newblock \href {https://arxiv.org/abs/2410.16153} {Pangea: A fully open multilingual multimodal llm for 39 languages}.
\newblock \emph{arXiv preprint arXiv:2410.16153}.

\bibitem[{Zhang et~al.(2024{\natexlab{a}})Zhang, Li, Liu, Lee, Gui, Fu, Feng, Liu, and Li}]{zhang2024llavanextvideo}
Yuanhan Zhang, Bo~Li, haotian Liu, Yong~jae Lee, Liangke Gui, Di~Fu, Jiashi Feng, Ziwei Liu, and Chunyuan Li. 2024{\natexlab{a}}.
\newblock \href {https://llava-vl.github.io/blog/2024-04-30-llava-next-video/} {Llava-next: A strong zero-shot video understanding model}.

\bibitem[{Zhang et~al.(2024{\natexlab{b}})Zhang, Wu, Li, Li, Ma, Liu, and Li}]{zhang2024videoinstructiontuningsynthetic}
Yuanhan Zhang, Jinming Wu, Wei Li, Bo~Li, Zejun Ma, Ziwei Liu, and Chunyuan Li. 2024{\natexlab{b}}.
\newblock \href {https://arxiv.org/abs/2410.02713} {Video instruction tuning with synthetic data}.
\newblock \emph{Preprint}, arXiv:2410.02713.

\end{thebibliography}

\appendix

\section{Survey Design and Results}
\label{appendix:survey}
Our survey was designed to explore how individuals who are blind or have low vision use AI models as visual assistants. The focus was on understanding the tasks they perform and the challenges they face. The survey combined multiple-choice and open-ended questions, allowing for both quantitative and qualitative insights. The responses are useful to help identify patterns and areas for improvement in AI models to better serve individuals with vision impairments. 

\subsection{Survey Construction}

The survey was carefully designed with input from individuals who are blind or have low vision to ensure it accurately reflected their experiences and needs. We implemented two feedback loops by engaging with blind participants during the design process, allowing us to refine questions and make sure the survey was accessible and relevant. 

\begin{figure*}
    \centering
    \includegraphics[width=0.7\linewidth]{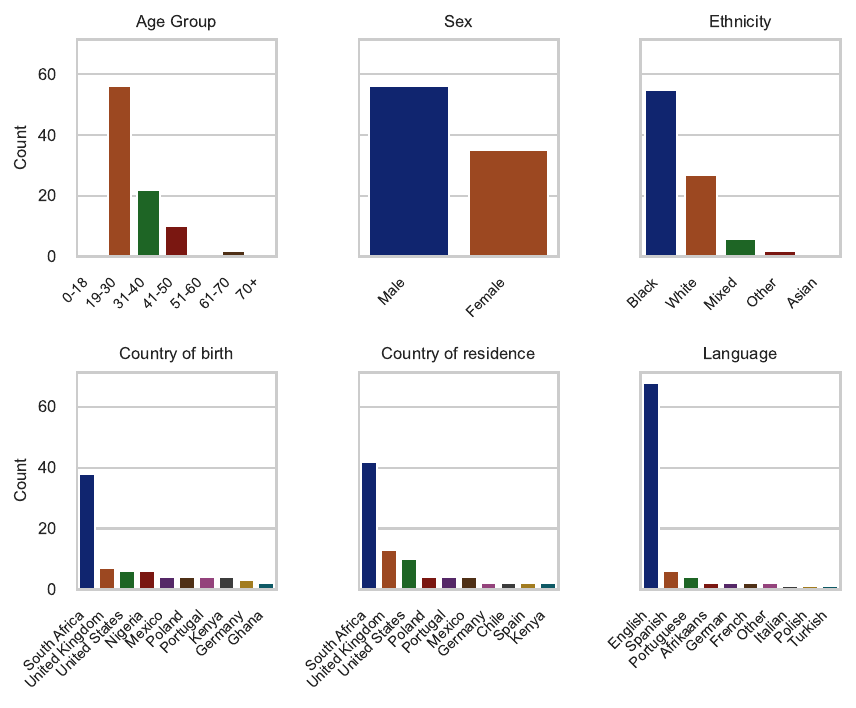}
    \caption{Age, Gender, and Ethnicity demographics extracted from Prolific after filtering the data to remove the "revoked\_consent" options.}
    \label{fig:demo1}
\end{figure*}

\paragraph{Demographics}
Participants were recruited via Prolific, and compensation was based on an average reward per hour (9 pounds) to ensure fair payment for their time. We asked for participants to be located across all countries available and for a fair distribution sample. We also added a screener for participants with no vision (found under Add Screeners<Health<No Vision). This resulted in 25,485 matching participants 'who have been active in the past 90 days'. We collected a total of 106 participants after filtering out some participants without visual impairments. Even though our survey was completely anonymous, Prolific provides some basic demographics for participants in a .csv format. We plotted some of the participant demographics after excluding the vision "yes" and "revoked\_consent" participations in \cref{fig:demo1}.

\subsection{Survey Sections and Results}

\paragraph{Introduction.}
Before beginning the survey, participants were briefed on its content and purpose: \textit{This survey is for individuals who are blind or have low vision and use AI models like ChatGPT or Gemini as visual assistants. Our goal is to understand the tasks they perform, the challenges they encounter, and their overall experiences with AI.}

We then obtained their consent, assuring them that their responses would remain completely anonymous—no email addresses or identifying information were requested. Participants were informed that the survey would contribute to a research project leading to a scientific publication and were encouraged to answer honestly and thoughtfully. Additionally, we provided contact details for both the student and their supervisor.

\begin{figure*}
    \centering
    \includegraphics[width=0.7\linewidth]{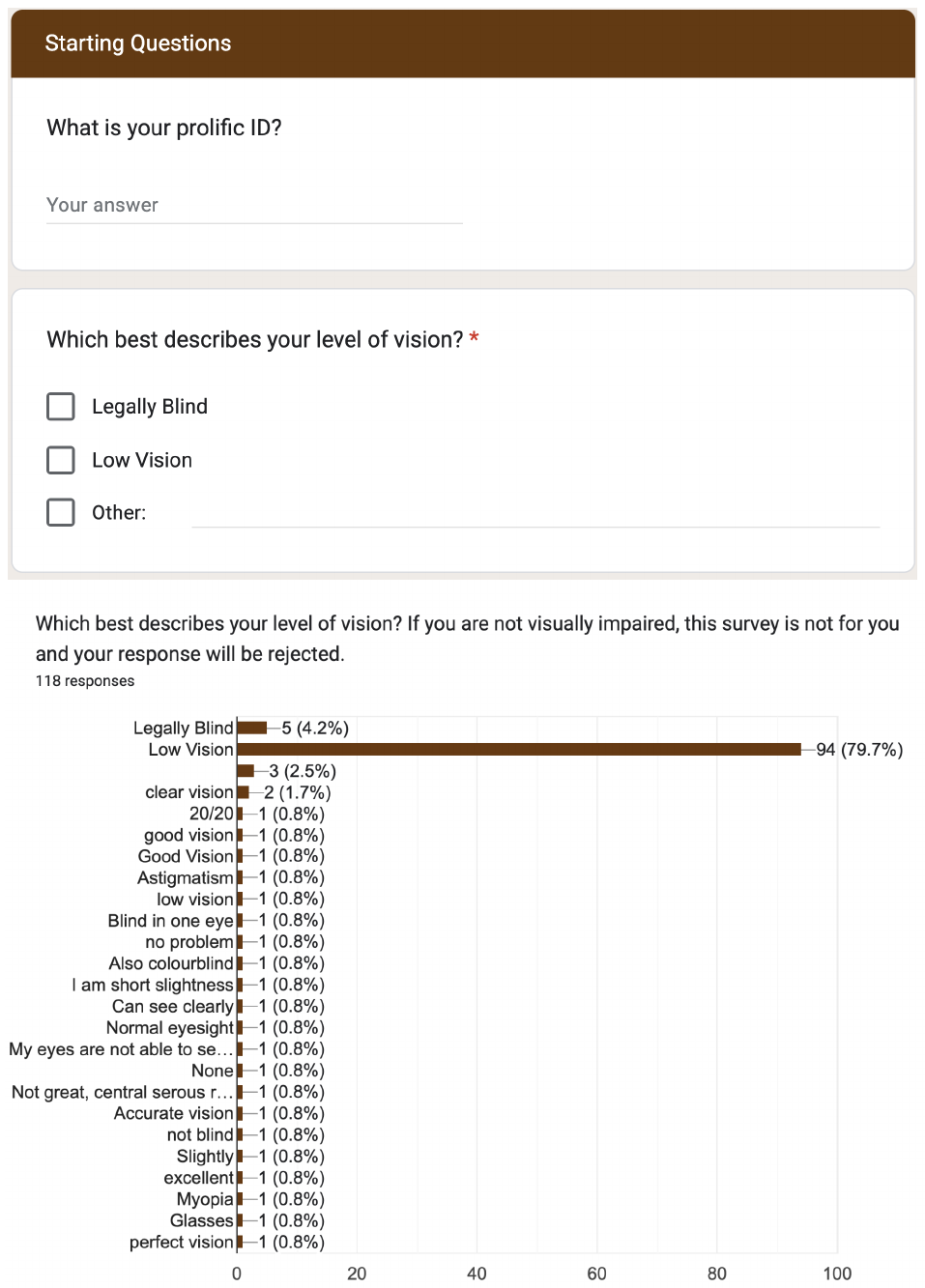}
    \caption{Phase 1: First section of the survey}
    \label{fig:firstsection}
\end{figure*}

At the beginning of the survey, we had some initial questions asking participants about their prolific ID, and we added an extra question to verify Prolific's screener is accurate and that we are going to get responses from visually impaired people. As shown in \cref{fig:firstsection}, there were actually some participants who did not have a visual impairment, so we had to filter their responses.

\begin{figure}[tb]
    \centering
    \includegraphics[width=0.5\columnwidth]{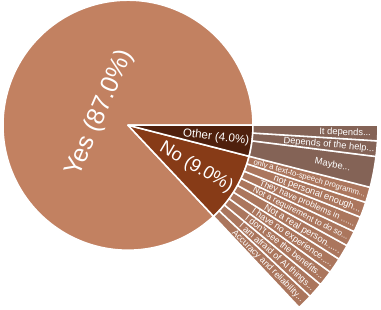}
    \caption{Responses on the potential adoption of AI models as visual assistants.}
    \label{fig:surveyresponses}
\end{figure}

\begin{figure*}
    \centering
    \includegraphics[width=0.7\linewidth]{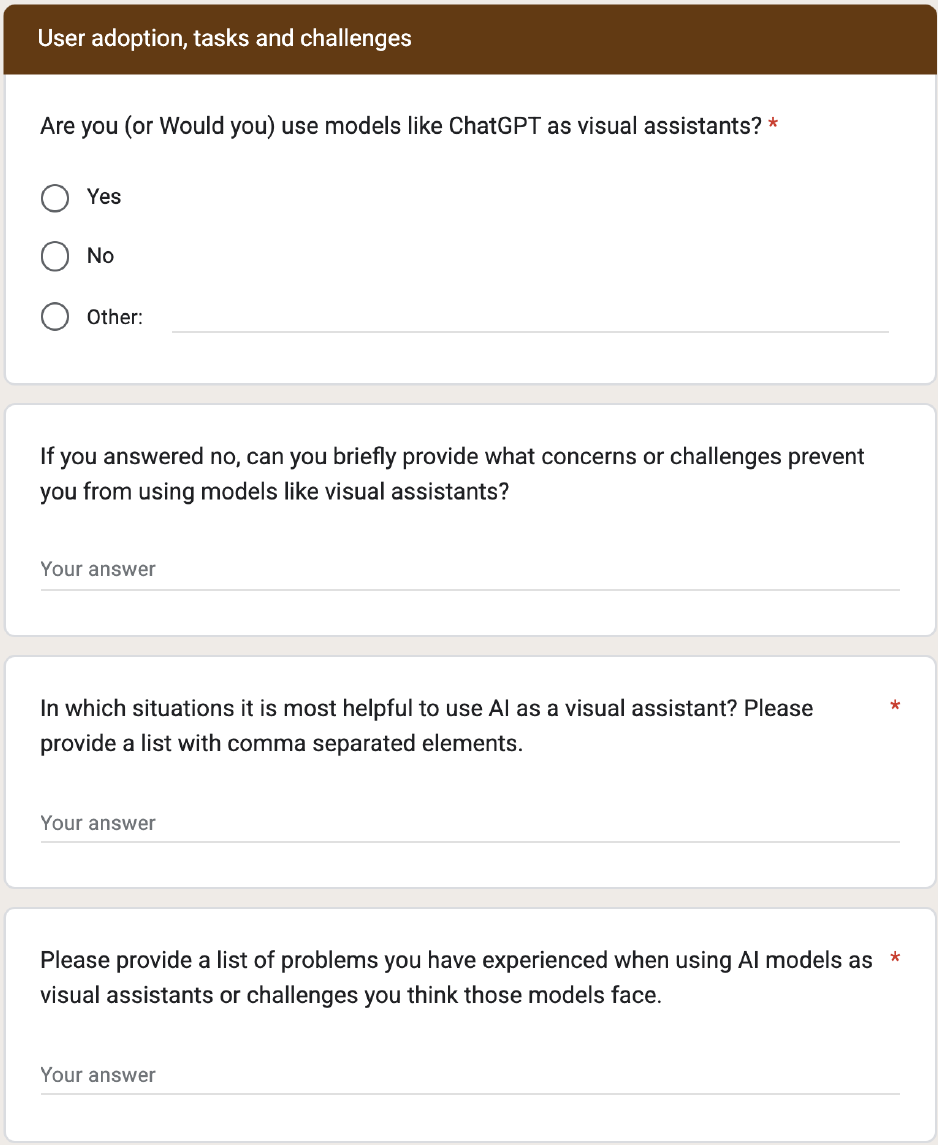}
    \caption{Phase 1: Second section of the survey}
    \label{fig:secondsection}
\end{figure*}

\paragraph{Phase 1: Open Ended Questions.}
The second section, as shown in \cref{fig:secondsection}, was about \emph{user adoption and tasks}, asking participants whether they currently use or would consider using Artificial Intelligence models as visual assistants. After filtering the responses from individuals without impairments, we visualize the results in \cref{fig:surveyresponses}. Most participants would use AI models as visual assistants, but there are some who are reluctant to use them. Participants were also asked to list situations where AI would be most helpful as a visual assistant, providing their responses as comma-separated elements. We left the question open to gather insights into the settings and tasks participants perform using AI models. The third question asked participants about problems and challenges they have experienced when using AI models as visual assistants. 

We conducted an iterative thematic analysis to better understand participants' perceptions of the open-ended questions, following \citep{researchtools}. Two authors reviewed and coded all responses into thematic categories. They then met with the research team to compare and finalize the themes. For the user tasks questions, we tried to use keywords from the responses to stay closer to the original task; for the challenges question, we tried to group the concerns under more generic themes. 

\begin{figure}[tb]
    \centering
    \includegraphics[width=0.9\columnwidth]{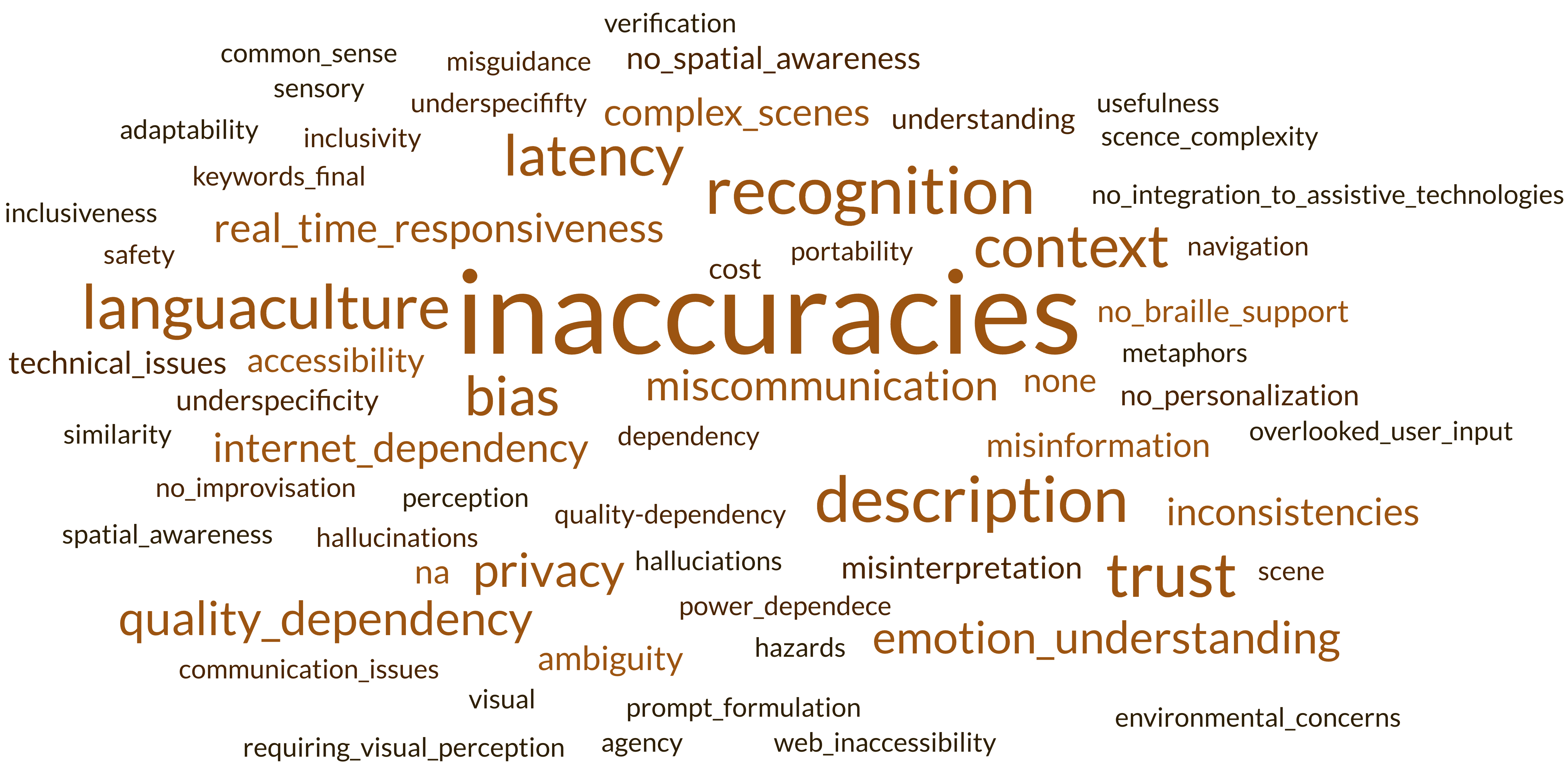}
    \caption{Visualizing all the concerns listed in our survey under the challenges open-ended question.}
    \label{fig:survey_wordcloyd}
\end{figure}

In ~\cref{fig:survey_wordcloyd} we present a wordcloud of term frequencies of extracted themes from the responses regarding AI problems and challenges. We also present further details of the most recurrent themes along with definitions and examples that justify their grouping in \cref{tab:survey_challenges_qualit}.

\begin{table*}[ht]
    \centering

    \small
    \begin{tabular}{p{3cm} p{5cm} p{7cm}}
    
        \toprule
        \textbf{Theme} & \textbf{Description} & \textbf{Example} \\
        \midrule
        \rowcolor{beige}

        \textbf{Inaccuracies} & Whether the model provides accurate predictions & 
        \textit{Accuracy is one issue whether this is about objects or faces},
        \textit{Innacurate object detection}; \textit{gives wrong directions}; \textit{can provide inaccurate information} \\
        
        \textbf{Context} & Ability to interpret information based on surrounding factors, background knowledge, and situational cues & \textit{problematic contextual understanding; inadequate localization and navigation; limited scene understanding; AI may not grasp complex spatial relationships or context; difficulty in recognizing context;	limited understanding of social cues} \\
        
        \rowcolor{beige}

        \textbf{Recognition} & Whether the model can recognize objects, faces, characters & 
        \textit{"Inaccurate Object Recognition; Object recognition errors"; difficulty recognizing elements in dimly lit environments; recognizing small or ambiguous objects or text; limitations in recognizing facial expressions; error in recognizing text in blurred or obstructed images} \\

        \textbf{Description} & Whether the model can describe objects or scenes (from image or video) & 
        \textit{"Difficulty describing complex scenes with multiple objects; Bad photo descriptions; inability to describe nuanced scenes; inability to describe subjective or abstract contexts and challenges in distinguishing similar objects"} \\
        \rowcolor{beige}

        \textbf{Languaculture} & Difficulties in effectively using or understanding language in multilingual or multicultural settings & \textit{Speaking in my native language is not natural; Does not speak my language; They cannot identify culture-specific photographs} \\
 
        \textbf{Privacy} & Whether data is kept private and stored securely & 
        \textit{Constant image and audio processing could raise security issues; Privacy concerns when analyzing personal images or surroundings} \\
               \rowcolor{beige}
        \textbf{Miscommunication} & Communication barriers and misinterpretation of user input or intent & 
        \textit{They misunderstand what I mean; I dont know how to describe something to an AI model or how to get a correct response; Not effectively understanding my need or description of the question } \\

        \textbf{Quality \newline Dependency} & Reliance on high-quality training data for accurate outputs & 
        \textit{Dependency on high-quality data; Problems with blurry images that are too colorful or lower quality; Weather and lighting conditions and poor image quality can reduce accuracy} \\
        \rowcolor{beige}
        \textbf{Latency} & The delay in processing or response time & 
        \textit{Speed of process is slow; Delays in processing can affect real-time assistance; Slow responses; No fast natural human-like answers} \\

        \textbf{Trust} & The belief or confidence in the reliability, and truth of the model outputs  & 
        \textit{I cannot trust it with confidence; No trust in description of images; There is no trust between users and technology; Detailed information cannot be trusted} \\
        
        \bottomrule
    \end{tabular}
    \caption{Themes found in phase 1 question about problems and challenges participants have experienced when using AI models as visual assistants.}
    \label{tab:survey_challenges_qualit}
\end{table*}

\begin{figure*}
    \centering
    \includegraphics[width=0.7\linewidth]{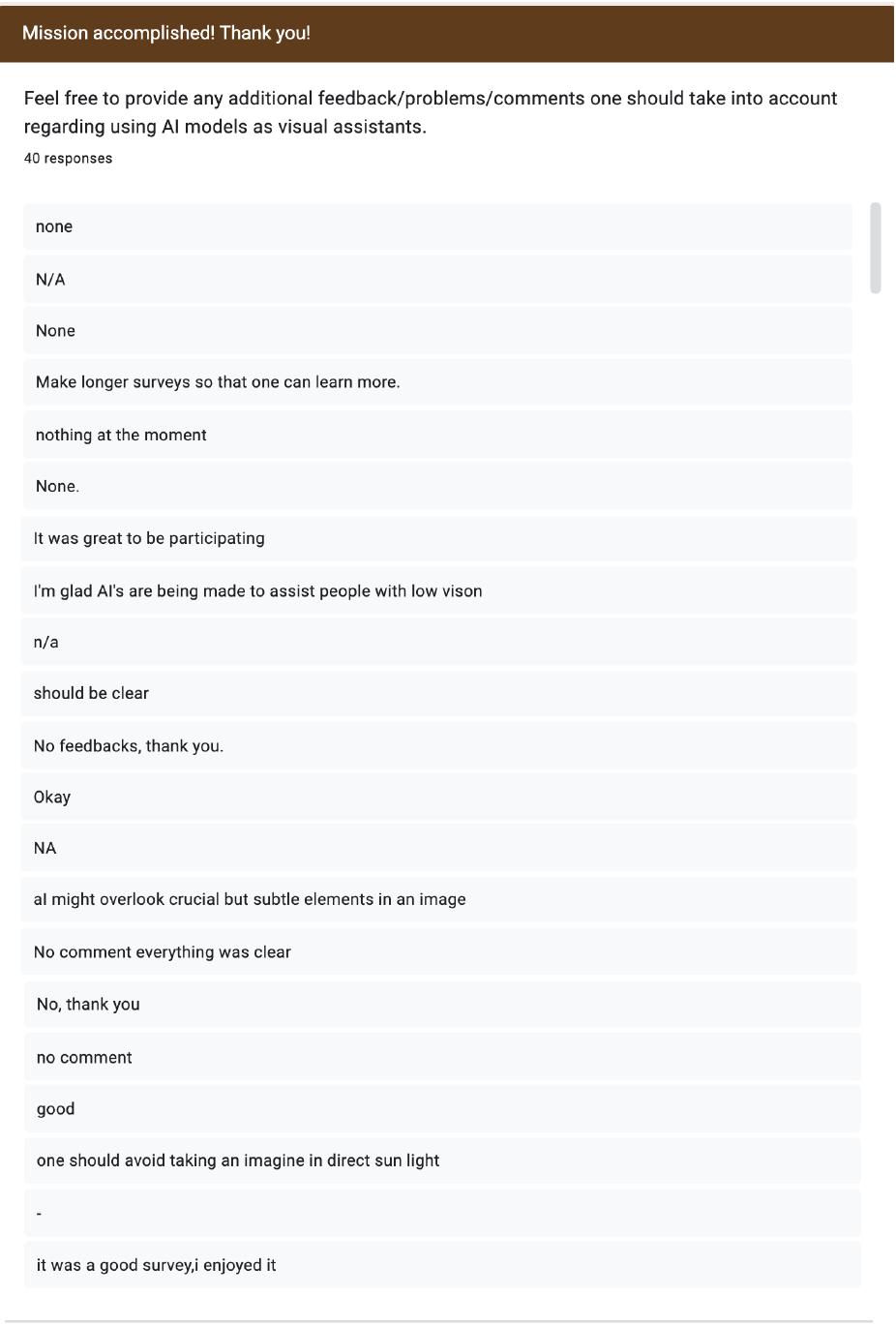}
    \caption{Phase 1: Last question before we direct the participants to the second phase.}
    \label{fig:lastsection}
\end{figure*}

The last section, as shown in \cref{fig:lastsection}, was optional and asked for any additional feedback or comments, but we only collected 40 responses, and most of them had no new insights.

\begin{figure}[tb]
    \centering
    \includegraphics[width=0.9\columnwidth]{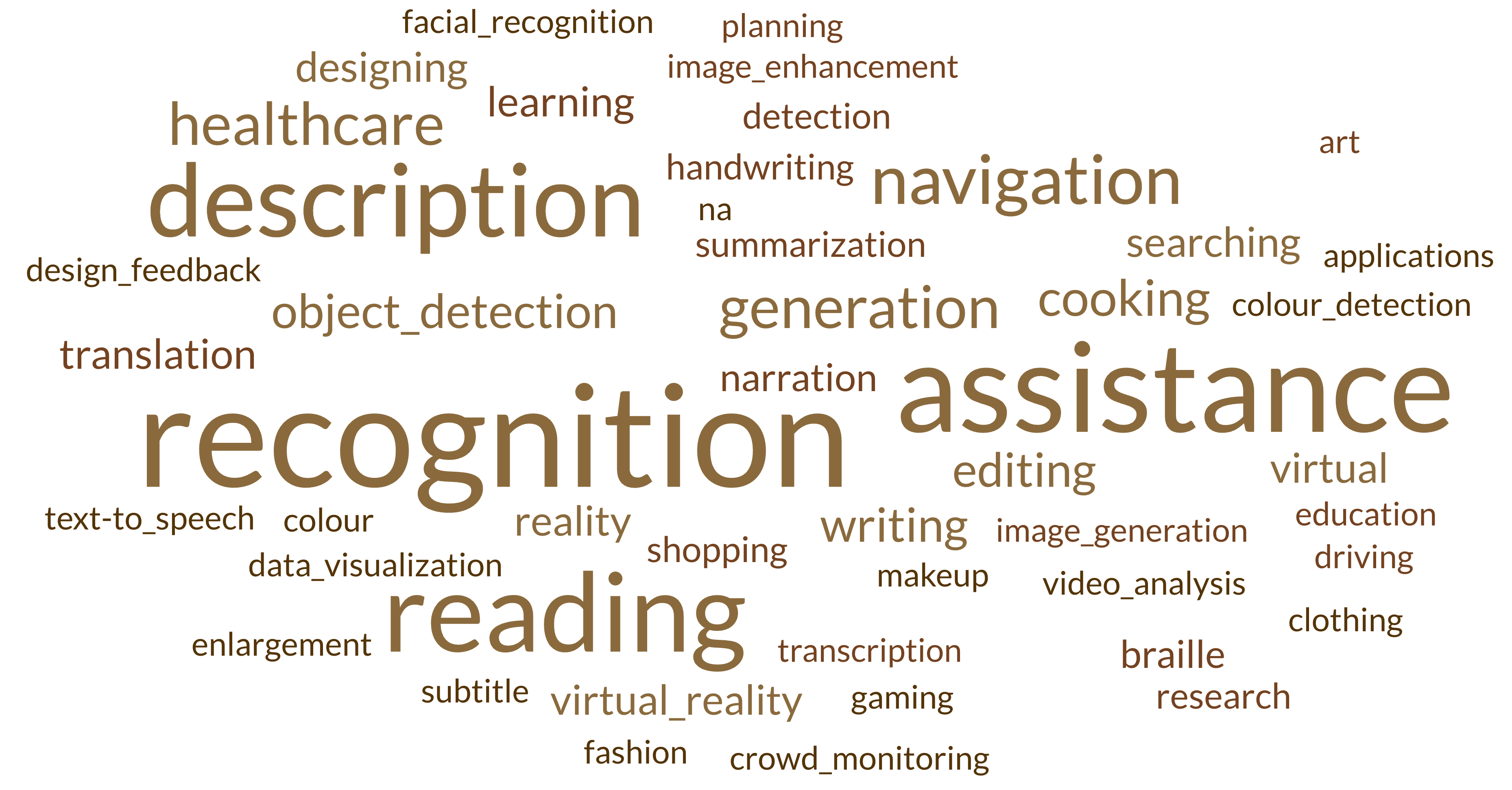}
    \caption{Visualizing all the user cases listed in our survey under the tasks open-ended question.}
    \label{fig:survey_wordcloyd2}
\end{figure}

\paragraph{Phase 2: Likert Scale Questions.}

\begin{figure*}
    \centering
    \includegraphics[width=0.7\linewidth]{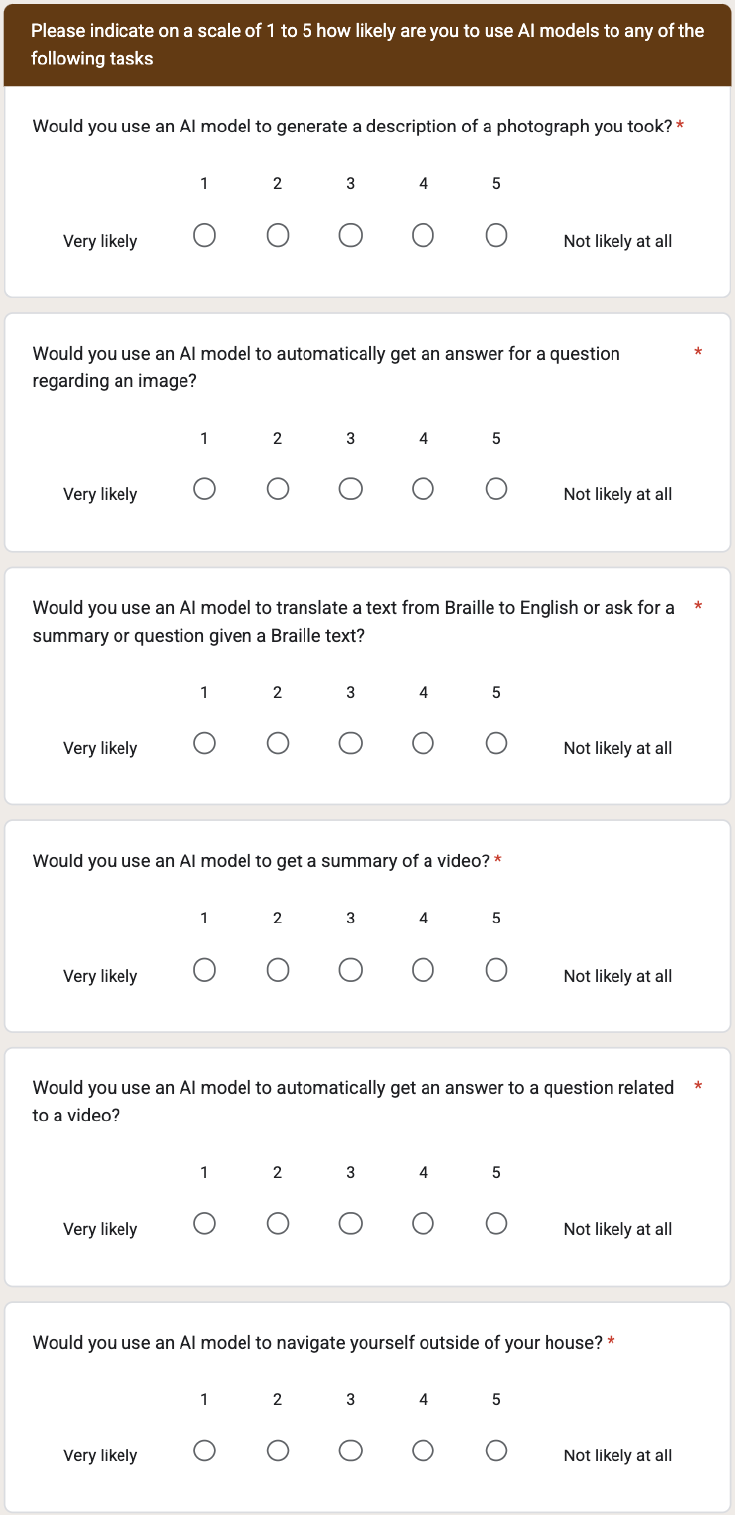}
    \caption{Phase 2: Asking participants to rate tasks that AI can be used for.}
    \label{fig:tasks-likert}
\end{figure*}

\begin{figure*}
    \centering
    \includegraphics[width=0.7\linewidth]{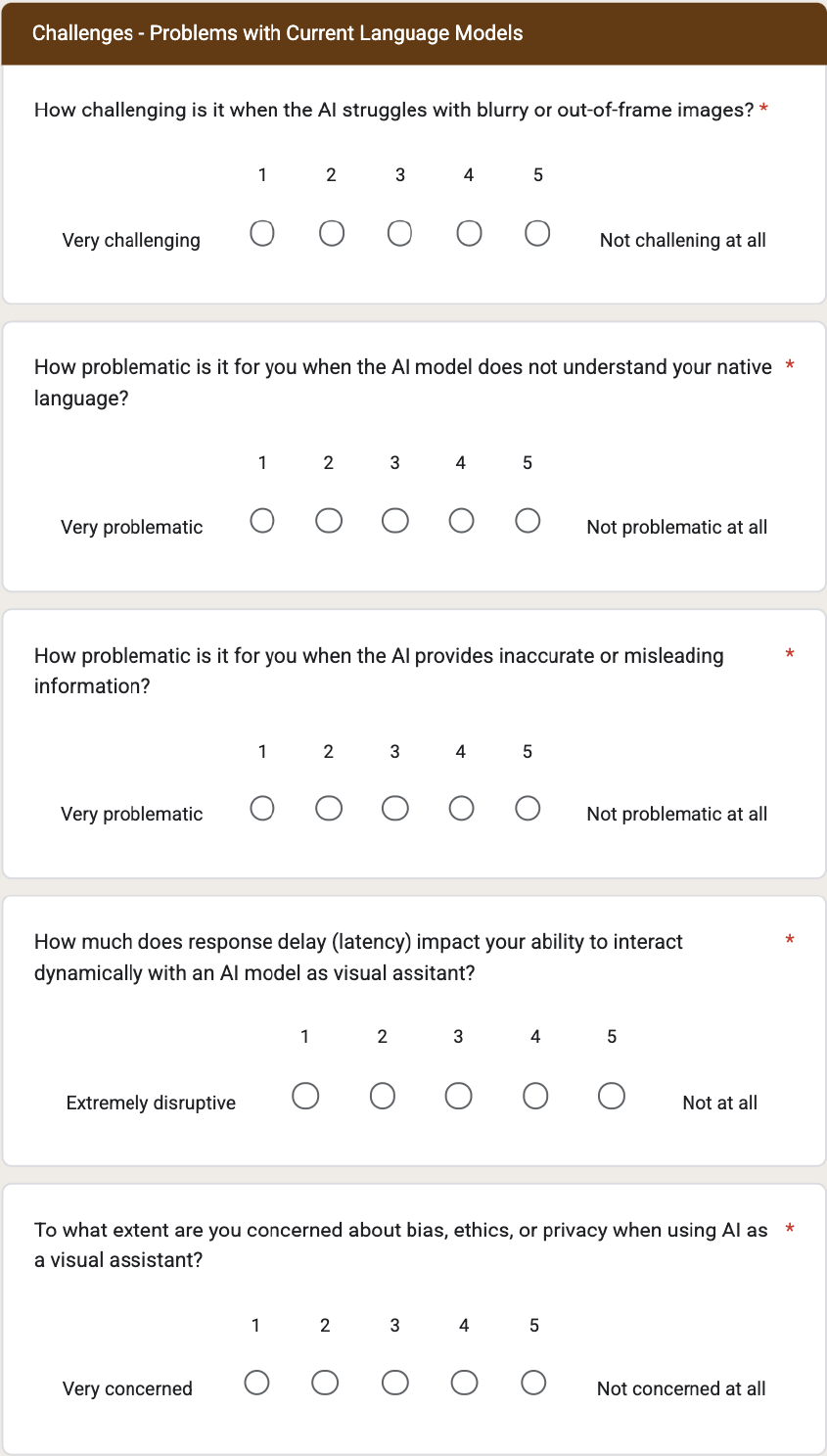}
    \caption{Phase 2: Asking participants to rate challenges they have encountered when using AI models.}
    \label{fig:problems-likert}
\end{figure*}

For the second phase of the survey, we asked the same participants after they provided their open-ended question answers to rate specific tasks and challenges. The task was to indicate on a scale of 1 to 5 how likely they are to use AI models for any of the following tasks: Image Captioning, Image Question Answering (IQA), Braille support, Video Question Answering (VQA), and Navigation. The exact phrasing of the questions can be seen in \cref{fig:tasks-likert}. We then asked them to indicate how problematic their shortcomings are related to image quality, language barriers, misinformation, latency, and bias. These categories were chosen based on the discussions we had with visually impaired users in the survey design phase. The exact phrasing of the questions can be seen in \cref{fig:problems-likert}.

\begin{figure*}[ht]
    \centering
    \includegraphics[width=0.7\linewidth]{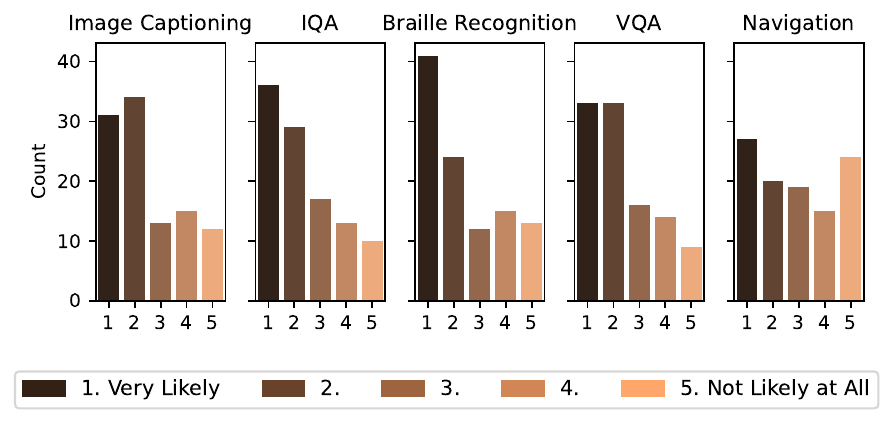}
    \caption{Likert scale responses for AI usage scenarios. The x-axis represents user responses ranging from 1 (Very Likely) to 5 (Not Likely at All), while the y-axis shows the count of responses for each question. }
    \label{fig:result-tasks-likert}
\end{figure*}

\begin{figure*}
    \centering
    \includegraphics[width=0.8\linewidth]{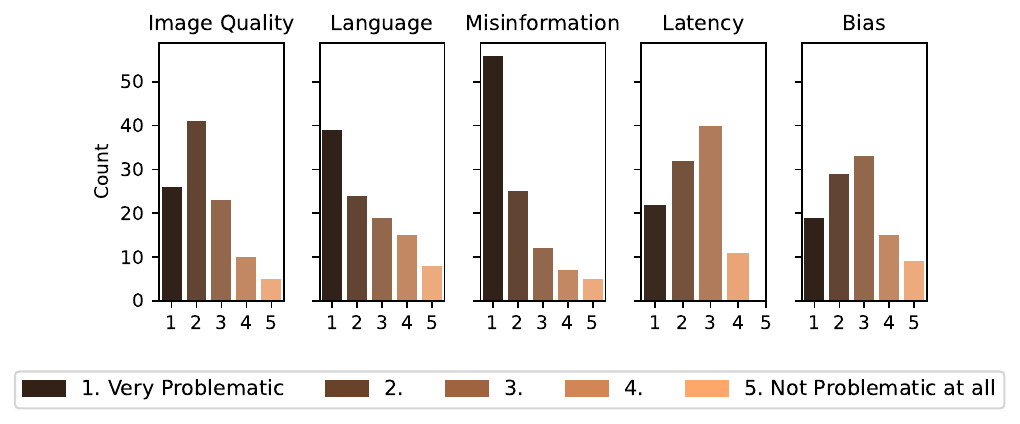}
    \caption{Likert scale responses for AI challenges. The x-axis represents user responses ranging from 1 (Very Problematic) to 5 (Not Problematic at All), while the y-axis shows the count of responses for each question. }
    \label{fig:result-problems-likert}
\end{figure*}

The results from Phase 2, presented in \cref{fig:result-tasks-likert} and \cref{fig:result-problems-likert}, indicate a growing adoption of AI models for tasks such as image captioning, question answering, and Braille recognition. However, opinions on using AI for navigation are more varied, with responses distributed across all possible values, suggesting that participants are not certain about using AI models as navigators.

Regarding challenges, misinformation appears to be the most common issue faced by participants, followed by language-related difficulties. Image quality also poses a problem, potentially affecting the reliability of AI-generated descriptions. While latency is a concern for many, a significant number of participants remain neutral on this issue. Notably, bias does not seem to be a major issue among respondents, indicating that, at least in their experience, AI models are perceived as relatively fair in their outputs.

We refrain from making broad generalizations and encourage readers to interpret these findings in the context of our sample size. Future research with larger and more diverse participant groups may provide further insights into these trends.


\section{Visual Question Answering}
\label{appendix:vqa}
Given the lack of multilingual visual question-answering datasets focused on visually impaired users \citep{karamolegkou-etal-2024-vision}, and the challenges in evaluating multilingual models for visual question answering \citep{pfeiffer-etal-2022-xgqa}, we decided to create a multilingual version of the existing VizWiz dataset \citep{gurari2018vizwiz}. The original dataset is under CC BY 4.0 license \footnote{\url{https://creativecommons.org/licenses/by/4.0/}}.

\subsection{Multilingual Dataset Construction}
\label{sec:multi}
\paragraph{Translation and Filtering}
We extend the VizWiz dataset to a multilingual setting by automatically translating the original questions and answers into 35 languages from the XM3600 benchmark \cite{ThapliyalCrossmodal2022}, shown in \cref{tab:languages}. 
We exclude Cuzco Quechua because it is not supported by most translation models.
We use the NLLB-Distilled-1.3B \cite{costa2022no} model to translate the question-answer pairs, given its strong performance and extensive language coverage.
Additionally, we sample for translation a stratified subset of 500 questions, utilizing the skill annotations to ensure representative coverage of different visual question-answering scenarios.

We follow the automatic translation process described by \citep{yue2024pangea}.
We generate multiple translations of each question-answer pair and employ backtranslation for filtering. 
Specifically, we keep the translation whose backtranslation to English has the highest BLEU score \cite{papineni2002bleu} with the original input as reference.

\paragraph{Evaluation}
For evaluation, we follow the VizWiz framework which relies on multiple answer references to compute the model accuracy. 
We extend the answer preprocessing to include non-English punctuation symbols and additionally perform unicode normalization on both predicted and ground truth answers.

\begin{table}[tb]
\small
    \centering
    \begin{tabular}{lccc}
        \toprule
        \textbf{Language} & \textbf{ISO Code} & \textbf{Script} & \textbf{Resource} \\
        \midrule
        Arabic & ar & Arabic & High \\
        Bengali & bn & Bengali & Mid \\
        Czech & cs & Latin & High \\
        Danish & da & Latin & Mid \\
        German & de & Latin & High \\
        Greek & el & Greek & Mid \\
        English & en & Latin & High \\
        Spanish & es & Latin & High \\
        Persian & fa & Arabic & High \\
        Finnish & fi & Latin & High \\
        Filipino & fil & Latin & Mid \\
        French & fr & Latin & High \\
        Hebrew & he & Hebrew & Mid \\
        Hindi & hi & Devanagari & High \\
        Croatian & hr & Latin & Mid \\
        Hungarian & hu & Latin & High \\
        Indonesian & id & Latin & Mid \\
        Italian & it & Latin & High \\
        Japanese & ja & Japanese & High \\
        Korean & ko & Hangul & High \\
        Māori & mi & Latin & Low \\
        Dutch & nl & Latin & High \\
        Norwegian & no & Latin & Low \\
        Polish & pl & Latin & High \\
        Portuguese & pt & Latin & High \\
        Romanian & ro & Latin & Mid \\
        Russian & ru & Cyrillic & High \\
        Swedish & sv & Latin & High \\
        Swahili & sw & Latin & Low \\
        Telugu & te & Telugu & Low \\
        Thai & th & Thai & Mid \\
        Turkish & tr & Latin & High \\
        Ukrainian & uk & Cyrillic & Mid \\
        Vietnamese & vi & Latin & High \\
        Chinese & zh & Han & High \\
        \bottomrule
    \end{tabular}
    \caption{Language, ISO-Codes, script, and resource levels.}
    \label{tab:languages}
\end{table}

\subsection{Human Evaluation of Automatic Translation}
\label{ref:app-data-valid}
To validate the quality of the automatically translated VizWiz QA data, we run a human evaluation. The evaluation focused on assessing the quality of machine-translated questions and answers while quantifying translation errors. At least 20 translated questions in random order were reviewed per language, each accompanied by 10 similar short answers. Evaluators assessed only the quality of the target language translation and provided relevant examples and comments in a spreadsheet tab corresponding to each language. 
The human evaluation guidelines are shown in \cref{fig:multilingual_guidelines}.

The study examined 7 languages selected based on the authors’ fluency. Each language was assessed using 220 data points, resulting in a total of 1,540 translated questions and answers. Of these, 257 were labeled as incorrect, yielding an average translation error rate of 16.28\%, representing the proportion of translations with noticeable errors. 

Most errors occurred because of improper translation of English brand names, which were mistakenly translated as generic words or altered (e.g., Gevalia Coffee, Diet Coke, Dr Pepper, Windows PC, LG, Mrs. Dash, Manwich). Additionally, there were issues with yes/no questions, where some languages produced incorrect responses such as double ‘yes, yes', ‘I don’t know', or ‘I am sorry' instead of a simple yes or no.

Some errors also resulted from problematic original answers that contained typos or ungrammatical phrases, such as ‘can diet' instead of ‘a can of Diet Coke' or ‘ginerale' instead of ‘ginger ale'. A notable case involved the number 321, which was mistranslated as a random sentence rather than being retained as a numeral.

Finally, two ambiguous words in the validation set—‘denomination' and ‘dressing'—posed challenges. Since the responses consisted of short, context-free answers, some models translated them with one interpretation, while others chose a different meaning, leading to inconsistencies across languages and the meaning of the correct response. We are going to release the dataset validation spreadsheet with the translation errors and the languages after the anonymity period.

\begin{figure*}[tb!]
    \centering
    \begin{tcolorbox}[title=Human Evaluation Instructions]

\underline{\textbf{\textsc{Multilingual Translation Evaluation}}} \\
\vspace{-2mm}

\textbf{Objective:} \\
Your task is to evaluate the translation quality of at least 20 machine-translated questions and their corresponding answers. 
(1 question has 10 similar short answers.) Focus only on the quality of the target language translation, not the accuracy of the question-answer content. Translations are provided in a JSON file, and results should be recorded in the spreadsheet tab labeled with your language name.

\vspace{2mm}
\textbf{Evaluation Process:}

\begin{enumerate}
    \item \textbf{Review Translations:}
    \begin{itemize}
        \item Read the translated answers for each question in the JSON file.
        \item If unsure about a translation, retrieve the original question using the image ID on this platform: \url{https://vizwiz.cs.colorado.edu/VizWiz_visualization/view_dataset.php}.
    \end{itemize}

    \item \textbf{Identify Errors:}
    Check for errors in each translated question-answer pair. For example, you can identify:
    \begin{itemize}
        \item \textbf{Grammatical Errors:} Incorrect grammar or sentence structure.
        \item \textbf{Lexical Errors:} Incorrect word choices or omissions.
        \item \textbf{Formatting Errors:} Issues with punctuation or capitalization.
    \end{itemize}

    \item \textbf{Assign Error Severity:}
    \begin{itemize}
        \item \textbf{Minor:} Small errors that do not impact the meaning.
        \item \textbf{Moderate:} Errors that partially affect clarity or meaning.
        \item \textbf{Severe:} Errors that significantly alter or obscure the meaning.
    \end{itemize}

    \item \textbf{Count Errors:}
    \begin{itemize}
        \item Track the total number of errors for each translated QA pair.
        \item Provide a severity score for each error identified.
    \end{itemize}
\end{enumerate}

\vspace{2mm}
\textbf{Report Results:} \\
Record your results in the spreadsheet using the provided structure:
\begin{verbatim}
ImageID, Error Count, Error Type, Comments, Examples
\end{verbatim}
If a QA pair has multiple error types, separate them with commas under \textit{Error Type}.
    \end{tcolorbox}
\caption{Guidelines for Multilingual Translation Evaluation.}
\label{fig:multilingual_guidelines}
\end{figure*}

\subsection{Further Results}
\label{ref:multilingual_results}
We report performance per language script in \cref{tab:vizwiz_vqa_script}, and per language in \cref{tab:multilingual_full}.
\begin{table}[th!]
    \centering
    \small
    \begin{tabular}{lccc}
    \toprule
    \textbf{Model} & \textbf{High} & \textbf{Mid} & \textbf{Low} \\\midrule
    Idefics3 & 24.8 & 20.8 & 21.7 \\
    InternVL2.5-MPO & 40.3 & 36.4 & 39.6 \\
    Llava-v1.6 & 41.9 & 37.7 & 43.3 \\
    Llama-3.2-Vision-Instruct & 29.8 & 27.6 & 33.4 \\
    Molmo & 28.2 & 28.5 & 32.6 \\
    MiniCPM-2.6 & 32.1 & 31.0 & 22.7 \\
    PaliGemma & 19.5 & 13.7 & 11.2 \\
    Phi-3-Vision-Instruct & 36.9 & 36.3 & 34.6 \\
    Qwen2-VL-Instruct & 44.8 & 44.9 & 42.9 \\  
    \bottomrule
    \end{tabular}
    \caption{Accuracy on multilingual VizWiz grouped based on the language characterization as High, Mid, and Low resource.}
    \label{tab:vizwiz_vqa_resource}
\end{table}

\begin{table*}[th!]
    \centering
    \scriptsize
    \addtolength{\tabcolsep}{-0.25em}
    \begin{tabular}{l cccccccccccc}\toprule
    \textbf{Model} & \textbf{Latin} & \textbf{Han} & \textbf{Japanese} & \textbf{Hangul} & \textbf{Cyrillic} & \textbf{Arabic} & \textbf{Devanagari} & \textbf{Hebrew} & \textbf{Thai} & \textbf{Telugu} & \textbf{Greek} & \textbf{Bengali} \\\midrule
        Idefics3 & 26.1 & 11.2 & 8.9 & 19.2 & 17.5 & 23.7 & 27.0 & 25.9 & 11.5 & 13.4 & 20.8 & 20.3 \\
        InternVL2.5-MPO & 40.5 & 44.1 & 39.7 & 25.9 & 27.2 & 41.3 & 35.6 & 43.6 & 42.0 & 39.6 & 39.3 & 29.3 \\
        Llava-v1.6 & 42.9 & 43.2 & 34.3 & 42.0 & 24.9 & 42.5 & 41.2 & 44.0 & 42.6 & 42.9 & 41.7 & 19.0 \\
        Llama-3.2-Vision-Instruct & 31.1 & 35.9 & 23.5 & 20.7 & 23.2 & 30.7 & 16.9 & 34.3 & 37.9 & 30.9 & 26.8 & 18.0 \\
        Molmo & 31.8 & 23.6 & 6.8 & 32.1 & 25.5 & 22.0 & 12.8 & 42.3 & 19.4 & 30.1 & 33.2 & 12.5 \\
        MiniCPM-2.6 & 34.0 & 43.4 & 32.2 & 21.4 & 24.4 & 22.1 & 11.0 & 31.8 & 33.5 & 14.4 & 34.9 & 10.3 \\
        PaliGemma & 18.4 & 15.5 & 26.9 & 22.9 & 13.3 & 8.9 & 9.9 & 10.8 & 13.1 & 20.5 & 9.9 & 12.5 \\
        Phi-3-Vision-Instruct & 38.7 & 33.7 & 31.5 & 34.3 & 26.0 & 33.0 & 30.4 & 37.7 & 37.7 & 30.1 & 35.5 & 35.5 \\
        Qwen2-VL-Instruct & 45.5 & 42.7 & 36.9 & 44.6 & 43.0 & 43.9 & 44.2 & 42.2 & 46.3 & 42.9 & 46.4 & 42.1 \\
  \bottomrule
    \end{tabular}
    \caption{Accuracy on multilingual VizWiz per script.}
    \label{tab:vizwiz_vqa_script}
\end{table*}

\begin{table*}[th!]
    \centering
    \scriptsize
    \addtolength{\tabcolsep}{-0.4em}
    \begin{tabular}{l cccccccccccccccccc}\toprule
    \textbf{Model} & \textbf{ar} & \textbf{bn} & \textbf{cs} & \textbf{da} & \textbf{de} & \textbf{el} & \textbf{en} & \textbf{es} & \textbf{fa} & \textbf{fi} & \textbf{fil} & \textbf{fr} & \textbf{he} & \textbf{hi} & \textbf{hr} & \textbf{hu} & \textbf{id} & \textbf{it} \\\midrule
    Idefics3 & 25.7 & 20.3 & 22.4 & 27.8 & 37.4 & 20.8 & 50.4 & 26.8 & 21.7 & 25.6 & 14.1 & 22.4 & 25.9 & 27.0 & 21.8 & 19.7 & 30.9 & 31.9 \\
    InternVL2.5-MPO & 42.9 & 29.3 & 38.9 & 36.3 & 43.3 & 39.3 & 60.2 & 37.8 & 39.8 & 41.0 & 43.3 & 40.7 & 43.6 & 35.6 & 41.8 & 32.4 & 35.8 & 41.6 \\
    Llava-v1.6 & 43.6 & 19.0 & 40.8 & 43.3 & 46.5 & 41.7 & 60.1 & 42.0 & 41.4 & 40.7 & 43.7 & 44.0 & 44.0 & 41.2 & 39.4 & 41.8 & 45.1 & 44.7 \\
    Llama-3.2-Vision-Instruct & 41.3 & 18.0 & 32.0 & 20.8 & 33.4 & 26.8 & 45.1 & 35.8 & 20.1 & 33.3 & 15.7 & 32.2 & 34.3 & 16.9 & 28.3 & 23.7 & 37.2 & 36.9 \\
    Molmo & 38.9 & 12.5 & 24.6 & 27.2 & 23.6 & 33.2 & 43.6 & 31.7 & 5.2 & 25.4 & 38.4 & 32.6 & 42.3 & 12.8 & 29.7 & 32.4 & 40.8 & 31.6 \\
    MiniCPM-2.6 & 39.8 & 37.6 & 16.9 & 3.6 & 8.3 & 36.8 & 30.8 & 44.0 & 32.4 & 34.1 & 30.3 & 39.5 & 55.3 & 70.2 & 34.3 & 40.8 & 5.0 & 6.0 \\
    PaliGemma & 5.6 & 12.5 & 12.3 & 13.5 & 26.2 & 9.9 & 78.3 & 10.6 & 12.2 & 26.0 & 16.7 & 4.7 & 10.8 & 9.9 & 9.9 & 16.6 & 19.4 & 12.0 \\
    Phi-3-Vision-Instruct & 33.6 & 35.5 & 36.6 & 38.3 & 37.5 & 35.5 & 51.0 & 37.4 & 32.5 & 36.9 & 41.0 & 36.3 & 37.7 & 30.4 & 38.2 & 39.6 & 41.1 & 39.4 \\
    Qwen2-VL-Instruct & 45.3 & 42.1 & 41.6 & 47.2 & 48.2 & 46.4 & 68.6 & 43.7 & 42.5 & 45.6 & 47.8 & 44.8 & 42.2 & 44.2 & 44.7 & 35.4 & 48.9 & 44.7 \\  
        \toprule
       & \textbf{ja} & \textbf{ko} & \textbf{mi} & \textbf{nl} & \textbf{no} & \textbf{pl} & \textbf{pt} & \textbf{ro} & \textbf{ru} & \textbf{sv} & \textbf{sw} & \textbf{te} & \textbf{th} & \textbf{tr} & \textbf{uk} & \textbf{vi} & \textbf{zh} \\\midrule
        Idefics3 & 8.9 & 19.2 & 28.9 & 27.8 & 25.8 & 25.3 & 26.1 & 22.3 & 22.7 & 21.8 & 18.8 & 13.4 & 11.5 & 17.2 & 12.4 & 28.8 & 11.2 \\
        InternVL2.5-MPO & 39.7 & 25.9 & 43.3 & 44.4 & 40.5 & 40.5 & 40.2 & 42.6 & 44.8 & 30.3 & 34.8 & 39.6 & 42.0 & 40.0 & 9.6 & 41.8 & 44.1 \\
        Llava-v1.6 & 34.3 & 42.0 & 43.0 & 45.3 & 44.0 & 41.4 & 41.6 & 40.8 & 32.3 & 36.5 & 43.2 & 42.9 & 42.6 & 35.0 & 17.4 & 40.6 & 43.2 \\
        Llama-3.2-Vision-Instruct & 23.5 & 20.7 & 32.2 & 30.8 & 39.9 & 26.2 & 39.3 & 32.4 & 21.3 & 21.9 & 30.6 & 30.9 & 37.9 & 25.6 & 25.0 & 30.7 & 35.9 \\
        Molmo & 6.8 & 32.1 & 39.2 & 28.1 & 33.0 & 28.5 & 38.8 & 30.8 & 40.2 & 37.7 & 28.0 & 30.1 & 19.4 & 17.4 & 10.8 & 36.0 & 23.6 \\
        MiniCPM-2.6 & 36.3 & 20.4 & 38.3 & 27.5 & 36.5 & 37.0 & 35.8 & 27.7 & 16.7 & 5.3 & 34.7 & 38.3 & 21.5 & 33.7 & 41.0 & 42.0 & 33.8 \\
        PaliGemma & 26.9 & 22.9 & 1.2 & 41.2 & 12.7 & 12.1 & 18.2 & 21.2 & 16.9 & 19.2 & 10.5 & 20.5 & 13.1 & 13.8 & 9.7 & 9.3 & 15.5 \\
        Phi-3-Vision-Instruct & 31.5 & 34.3 & 39.2 & 36.5 & 38.5 & 38.3 & 34.5 & 43.5 & 37.5 & 43.8 & 30.5 & 30.1 & 37.7 & 42.0 & 14.5 & 31.7 & 33.7 \\
        Qwen2-VL-Instruct & 36.9 & 44.6 & 41.9 & 46.4 & 43.7 & 45.3 & 37.9 & 46.8 & 49.0 & 44.9 & 43.1 & 42.9 & 46.3 & 42.3 & 37.1 & 47.1 & 42.7 \\
    \bottomrule
    \end{tabular}
    \caption{Accuracy on multilingual VizWiz per language.}
    \label{tab:multilingual_full}
\end{table*}

\section{Optical Braille Recognition}\label{appendix:obr}

\paragraph{Dataset Creation} 
We generate rendered images of Braille text as summarized in \cref{sec:Braille_dataset_creation}.
We apply augmentations to the images from both transcription and cross-script QA tasks using the imgaug library \citep{imgaug}. 
More specifically, we use color, edge, geometric, contrast, and blur transformations families, where an image can be transformed with multiple of these augmentations at the same time.
For color, we select one of posterize, color quantization, and color temperature.
With regards to edge transformation, we either sharpen, emboss the image, or convert edges into black or white and overlay the resulted transformation with the original image.
For geometric transformations, we shear the image over the width or height or rotate the image.
Additionally, we scale pixel values by a fixed gamma constant.
Finally, we apply either gaussian, bilataral, motion or mean shift blur.
All augmentations are applied in random order.
The values for all of the parameters, along with scripts to reproduce the augmentations, are available \texttt{linked removed for review}.

\cref{tab:obr-inputs-outputs} illustrates examples of inputs-outputs for both tasks where the Braille text is rendered in images that have been augmented and the model needs to output plain English text.
Note that in all cases, the correct output cannot be inferred unless the model is able to read the Braille content from the image.

\paragraph{Training Logs \& Hyperparameters} \cref{tab:obr_hyperparams} illustrates the hyperapameters used to finetune Llama-3.2-Vision-Instruct on both tasks for Optical Braille Recognition.
Note that the LoRA adapters are applied to the key and value weight matrices in each transformer layer following the default implementation \citep{hu2022lora}.
We expect that applying the adapters to other linears can further improve performance. 
All experiments were conducted using 1xH100 GPU.
Training logs for all runs are available \texttt{linked removed for review}.

\begin{table}[tb]
    \centering
    \small
    \renewcommand{\arraystretch}{1.2}
        \begin{tabular}{@{}l c l@{}} 
            \toprule
            \textbf{Hyperparameter} & \textbf{Values}\\
            \midrule
            global batch size & 64\\
            LR & \{1e-5, 1e-4, 5e-4\} \\
            lr schedule & cosine decay \\
            LR warmup & 0.03\\
            number of epochs & 1\\
            optimizer & AdamW\\
            LoRA rank & \{32, 64, 128, 256\} \\
            LoRA alpha & \{16, 32, 64, 128, 256, 512\}\\
            \bottomrule
        \end{tabular}
    \caption{Hyperparameters during both finetuning on both sentence-level and paragraph-level tasks.}
    \label{tab:obr_hyperparams}
\end{table}

\section{Video Object Recognition and Question Answering}

\begin{figure}[tb]
    \centering
        \includegraphics[width=0.5\textwidth]{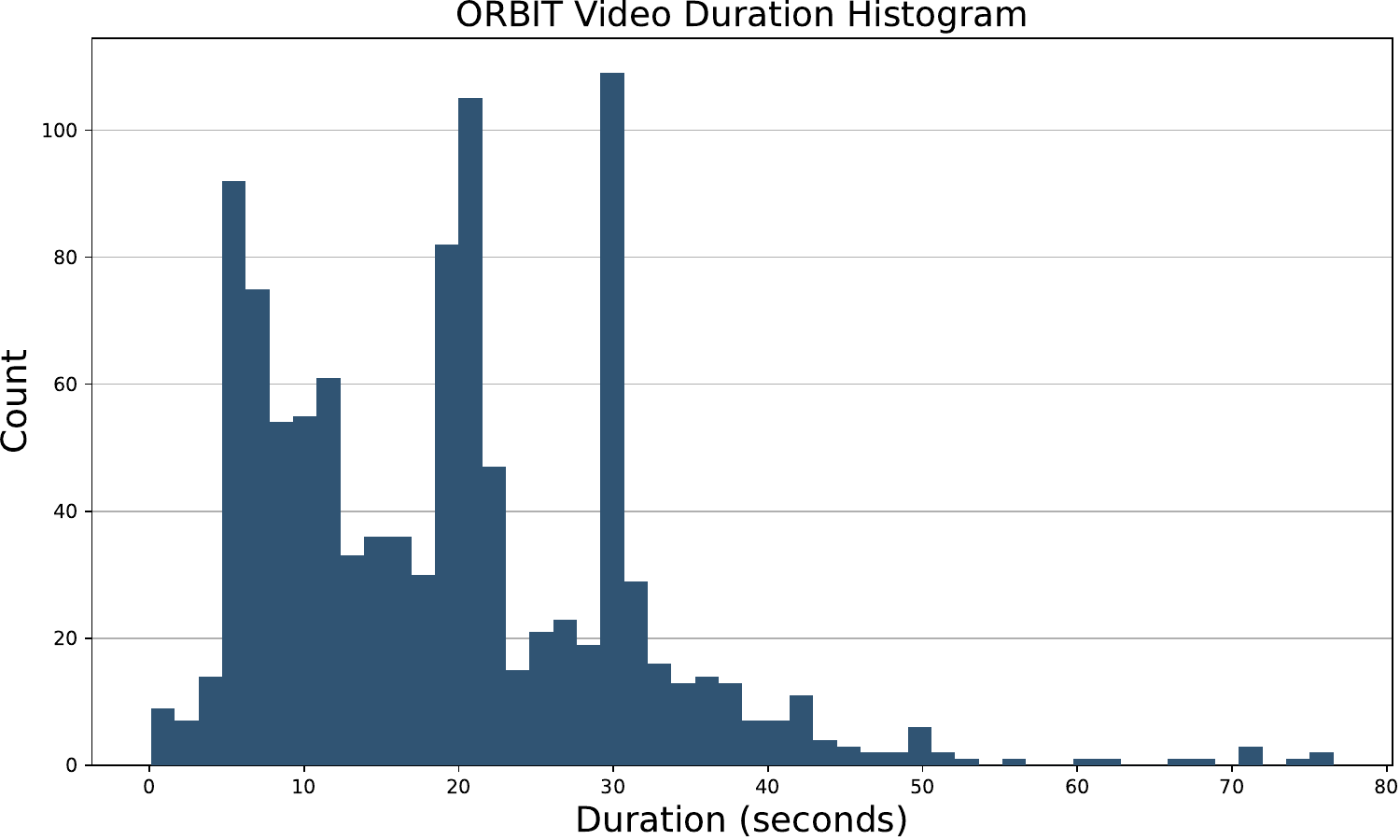}
        \caption{Video duration histogram.}
    \label{fig:orbit_hist}
\end{figure}

\begin{figure}[tb]
    \centering
        \includegraphics[width=0.5\textwidth]{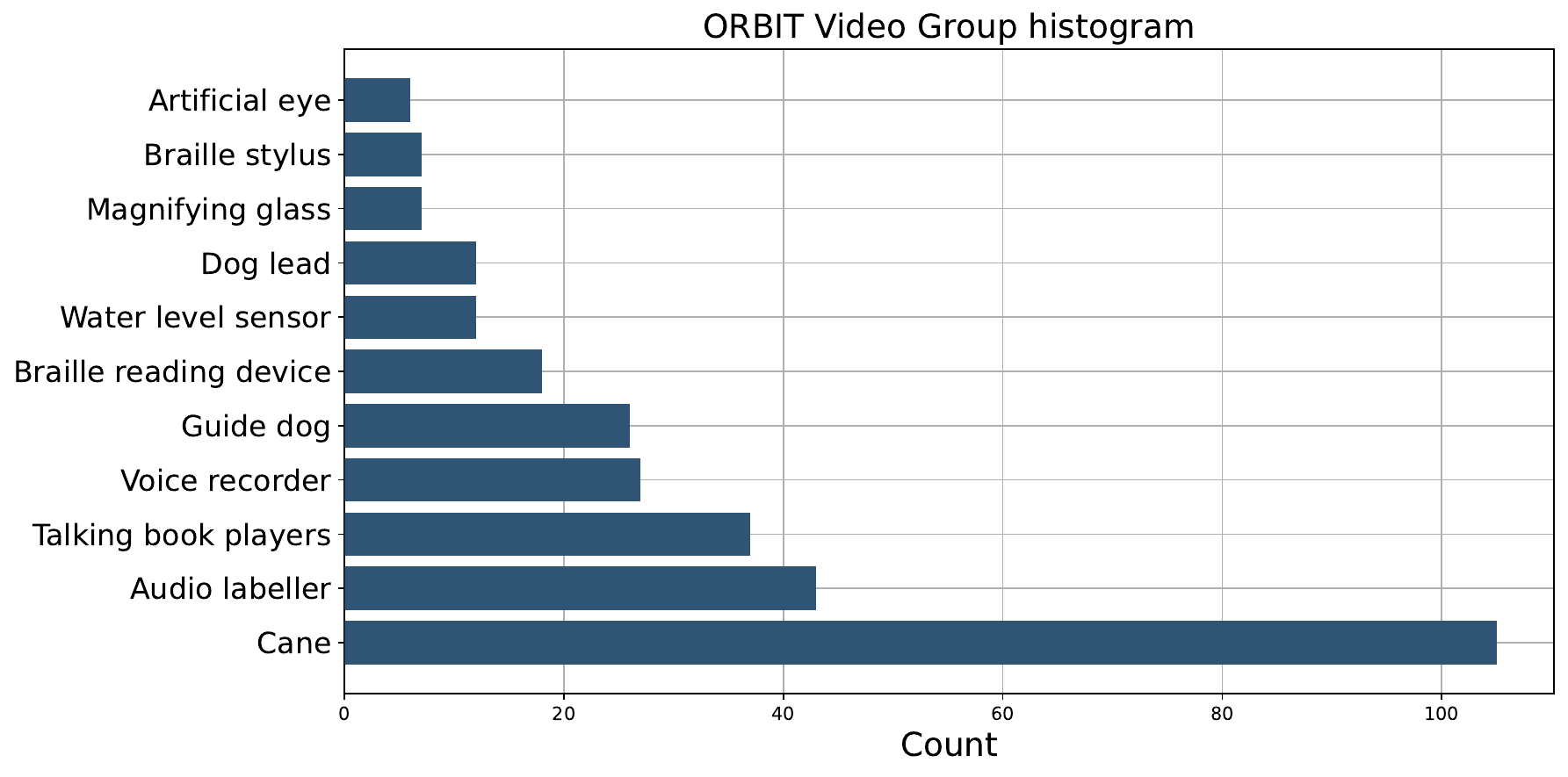}
        \caption{Number of videos per assistive category.}
    \label{fig:orbit_data_hist}
\end{figure}

\paragraph{ORBIT dataset}
ORBIT \cite{orbit} is a dataset of videos collected by people who are blind/low-vision, originally collected for few-shot object recognition. 
The dataset includes ``clean" videos, which show an object in isolation, and ``clutter" videos, which show the target object in the context of other items.
The target objects are labelled by the participants and grouped into clusters by the dataset authors. 
Videos are provided at 1080x1080 frame resolution and 30 frames per second.
We utilize 1069 video clips from 51 participants and 92 object clusters, with a median duration of 19.7 seconds (see \cref{fig:orbit_hist} for the video duration distribution).
The videos include household objects, which are general everyday objects (e.g., TV remote, house keys, wallet) and assistive items (e.g., Braille display, white cane, liquid level indicator), as illustrated in \cref{fig:orbit_data_hist}.

\begin{figure}[t]
    \centering
    \includegraphics[width=\linewidth]{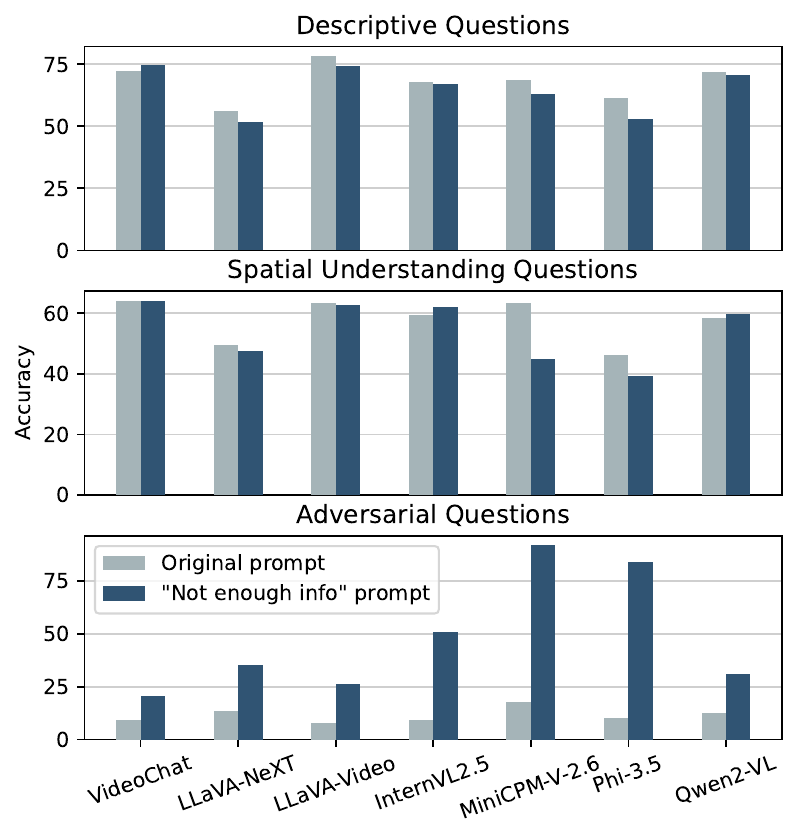}
    \caption{Accuracy on VideoQA when prompting the model to output ``Not enough information" as needed.}
    \label{fig:video_qa_prompt}
\end{figure}

\begin{figure}[t]
    \centering
    \includegraphics[width=\linewidth]{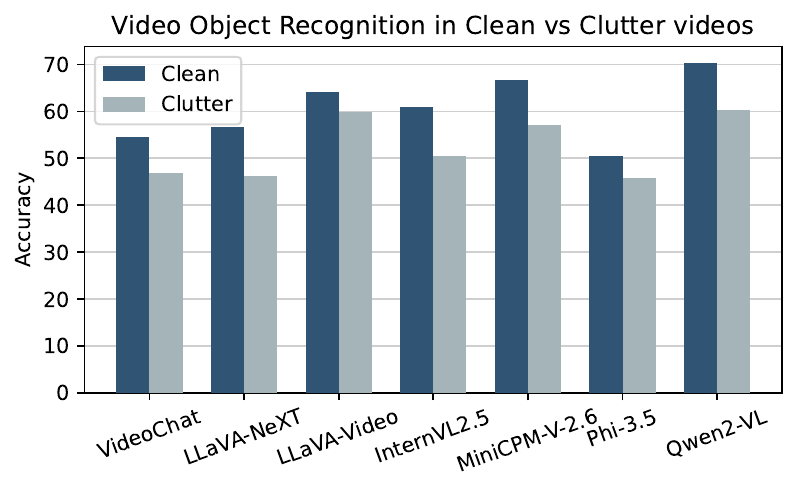}
    \caption{Accuracy in Video Object Recognition in Clean vs Clutter videos.}
    \label{fig:video_object_recognition_clean_vs_clutter}
\end{figure}

\subsection{Video Object Recognition Dataset Construction}
\label{appendix:vor}
For video object recognition, we use the dataset provided from \citep{orbit}. The dataset is under CC BY 4.0 license \footnote{\url{https://creativecommons.org/licenses/by/4.0/}}. We select 512 ``clean" and 514 ``clutter" videos through stratified sampling across object categories.
We convert the dataset into a question-answering format using a two-step semi-automatic process.
First, we prompt a language model to extract a representative keyword for each object cluster.
Second, based on these keywords and object labels, we generate an object recognition question for each group. 
The prompts used for the dataset creation are shown in \cref{fig:video_object_recognition_prompt}.
Finally, the generated questions are reviewed manually and adjusted if needed.

\begin{figure*}[ht]
    \centering
    \begin{tcolorbox}[title=Prompts for ORBIT Video Object Recognition Data Generation]

\underline{\textbf{\textsc{Keyword Extraction}}} \\
\vspace{-2mm}

You will be given a list of objects, and you have to answer with one short word or phrase that can be used to describe the group.
\vspace{1mm}

Examples\\
Objects: [watch, wrist watch, apple watch, apple wath, risk watch, my apple watch]\\
Answer: watch
\vspace{1mm}

Objects: [black small wallet, my purse, my wallet, ladies purse, money pouch, coin purse, wallet for bus pass cards and money, i d wallet, ipod in wallet, walletv, wallet, purse]\\
Answer: wallet
\vspace{1mm}

Objects: [orbit Braille reader and notetaker, orbit reader 20 Braille display, Braillepen slim Braille keyboard, Braille orbit reader, Braille note, my Braille displat]\\
Answer: Braille reading device
\vspace{1mm}

Generate the answer for the following objects:
\vspace{4mm}

\underline{\textbf{\textsc{QA Data Generation}}} \\
\vspace{-2mm}

You will be given a list of objects and a common label that describes the group. Your task is to generate a question that can be asked to identify an instance of this group in a video.
\vspace{1mm}

Examples:\\
Objects: [slippers, nike trainers, my shoes, boot, trainers, trainer shoe, slipper, my trainers, shoes, running shoes]\\
Group: shoes\\
Question: What type of clothing do you see in the video?
\vspace{1mm}

Objects: [orbit Braille reader and notetaker, orbit reader 20 Braille display, Braillepen slim Braille keyboard, Braille orbit reader, Braille note, my Braille displat]
Group: Braille reading device\\
Question: What kind of assistive device was there?
\vspace{1mm}

Objects: [black small wallet, my purse, my wallet, ladies purse, money pouch, coin purse, wallet for bus pass cards and money, i d wallet, ipod in wallet, walletv, wallet, purse]\\
Group: wallet\\
Question: What type of accessory appears in the video?
\vspace{1mm}

Generate the question for the following:
    \end{tcolorbox}
\caption{Prompts for the Video Object Recognition dataset.}
\label{fig:video_object_recognition_prompt}
\end{figure*}

\subsection{Video Question Answering Dataset Construction}
\label{appendix:qa}
Since there is currently no dataset with question answering for videos filmed by visually impaired users, we decided to curate such a dataset using videos from ORBIT \citep{orbit}. We use only ``clutter" videos that provide a more naturalistic setting.

We generate three types of questions:
1) Descriptive Questions, such as questions about color and number of objects, 2) Spatial Understanding, such as questions about the location or spatial relationship between objects and 3) Adversarial Questions which cannot be answered based on the information provided in the video.
To generate the questions, we used a manual approach where three of the authors of the paper followed the guidelines provided in \cref{fig:vqa_guidelines}. We create a total of 882 question-answer pairs (294 per question type). 

\begin{figure*}[tb]
    \centering
    \begin{tcolorbox}[title=Instructions for ORBIT Video QA Data Generation]

\underline{\textbf{\textsc{Annotation Guidelines for Video-Based Question Generation
}}} \\
\vspace{-2mm}

Step1: Video Access
Open this link containing short videos: \texttt{url} and the json file attached.
Watch 50 short video clips and generate 3 short questions + short answers about the clip.

Step2: Question Generation
Questions should be designed to help a Visually Impaired Person (VIP) understand and interact with their environment. They should be based on information that can be visibly inferred from the video. The focus should be on:

\begin{itemize}
    \item  Descriptive Questions (D)
These questions ask about the appearance, quantity, or basic attributes of objects.
Examples: "What is the colour of this item?", "How many X items do you see?", "What shape is this object?"

    \item  Spatial Understanding Questions (S)
These questions focus on the location and relationships between objects and people.
Examples: "What is next to this item?", "Where is item X?", "Is there an item Y next to item X?"

    \item  Adversarial Questions (A)
These questions ask for items or cues not present in the video.
Examples: "Is there an X in the image?", "Is there an X item next to the Y item?", "Is the colour of X item green?" 

The answer to this question is always: "Not enough information are depicted in the video to answer this question."
\end{itemize}

Step3: Answer Generation
Answers should be grounded in the information provided in the video. They should be short, clear, concise, and based on the video footage. For example Q: "How many X items are there", A: "four", Q: "Where is X item placed?" A: "inside a kitchen cabinet", Q: "Are there any mangoes next to the toy?" A: "No".

Step4: Write the video name id (eg. "P100--exercise bench--clutter-pan--P100--exercise-bench--clutter"), question, question\_type, and answer, in the json file. We provide some examples there for your guidance. 

    \end{tcolorbox}
\caption{Guidelines for Video Question Answering Data Generation}
\label{fig:vqa_guidelines}
\end{figure*}

\subsection{Evaluation Metric}
Given that only one label is available for each question, we adopt the LAVE metric \cite{manas2024improving} for evaluation. LAVE uses a language model judge to provide a rationale and a rating between 1-3. Ratings are then normalized in the range [0, 1].
We use Llama-3.3-70B-Instruct \cite{llama3modelcard} as the language model.

\begin{table*}[th!]
    \centering
    \small
    \resizebox{\textwidth}{!}{%
    \addtolength{\tabcolsep}{-0.2em}
    \begin{tabular}{l lccccc cc}
    \toprule
   \textbf{Model} & \textbf{Huggingface Tag} & \textbf{Param} & \textbf{Image} & \textbf{Video} & \textbf{Multilingual} & \textbf{Trained on}\\
    & & & & & \textbf{(\# Langs)} & \textbf{VizWiz}\\
    \midrule
    Idefics3 \citeyearpar{laurenccon2024building} & HuggingFaceM4/Idefics3-8B-Llama3 & 8B & \cmark & \xmark & \xmark & \xmark \\
    InternVL2.5-MPO \citeyearpar{wang2024enhancing} & OpenGVLab/InternVL2\_5-8B-MPO & 8B & \cmark & \cmark & \cmark (11) & \xmark\\ 
    LLaVA-NeXT-Video \citeyearpar{zhang2024llavanextvideo} & llava-hf/LLaVA-NeXT-Video-7B-hf & 7B & \xmark & \cmark & \xmark & \xmark \\
    LLaVA-Video \citeyearpar{zhang2024videoinstructiontuningsynthetic} & lmms-lab/LLaVA-Video-7B-Qwen2 & 7B & \xmark & \cmark & \xmark & \cmark$^*$ \\
    LlaVA-v1.6 \citeyearpar{liu2024improved} & llava-hf/llava-v1.6-mistral-7b-hf & 8B & \cmark & \xmark & \xmark & \xmark \\ 
    Llama-3.2-Vision-Instruct \citeyearpar{llama3.2} & meta-llama/Llama-3.2-11B-Vision-Instruct & 11B & \cmark & \xmark & \xmark & -- \\ 
    MiniCPM-V-2.6 \citeyearpar{yao2024minicpm} & openbmb/MiniCPM-V-2\_6 & 8B & \cmark & \xmark & \cmark (36) & \cmark\\ 
    Molmo \citeyearpar{deitke2024molmo} & allenai/Molmo-7B-D-0924 & 7B & \cmark & \xmark & \xmark & \xmark \\ 
    PaliGemma \citeyearpar{beyer2024paligemma} & google/paligemma-3b-mix-448  & 3B & \cmark & \xmark & \cmark (35) & \cmark\\ 
    Phi-3.5-Vision-Instruct \citeyearpar{abdin2024phi} & microsoft/Phi-3.5-vision-instruct & 4B & \cmark & \cmark & \cmark (--) & -- \\
    Qwen2-VL-Instruct \citeyearpar{wang2024qwen2} & Qwen/Qwen2-VL-7B-Instruct & 8B & \cmark & \cmark &  \cmark (--) & -- \\
    VideoChat-Flash \citeyearpar{li2024videochat} & OpenGVLab/VideoChat-Flash-Qwen2-7B\_res448 & 8B & \cmark & \cmark & \xmark & \xmark \\
    \bottomrule
    \end{tabular}}
\caption{Model Details. The model pool is limited to 1) open-source/weights models with 2) strong image or video understanding capabilities and 3) medium computational overhead. `--' is used when there is insufficient public information to determine the value. * VizWiz included in the image training phase before video instruction tuning.}
\label{tab:models}
\end{table*}

\subsection{Further Results}
\label{sec:videoqa-prompt_ablation}
Given the low performance on adversarial questions, we explore whether explicit prompting can mitigate this shortcoming. We modify the prompt to instruct models to respond with ``Not enough information" when the video content is insufficient to answer the question. 
As shown in \cref{fig:video_qa_prompt}, performance in Adversarial Questions consistently improves with the ``Not enough information" prompt. For most models, however, performance remains poor (at most 50\% accuracy) with minimal effect on other question types. This suggests that models continue to hallucinate answers frequently.
While performance increases drastically for MiniCPM-V-2.6 and Phi-3.5-Vision-Instruct, this comes at the cost of performance in other categories, as models tend to unnecessarily over-generate the ``Not enough information" response.
These results suggest that current prompting strategies alone cannot reliably prevent hallucination in video question answering—a critical safety concern for assistive applications.

\section{Models}
\label{appendix: models}
\cref{tab:models} reports the details for selected models, including the Huggingface tag used when accessing the model, the total number of model parameters, and whether models support image, video, and multilingual inputs. Additionally, we report whether VizWiz is included in the model's training data.
We find no evidence that any models are exposed to the ORBIT dataset.
Note that Paligemma is the only model that is not instruction fine-tuned, which is why we exclude it from zero-shot results in optical Braille recognition (\cref{tab:obr_zeroshot}).

\section{Examples}
\cref{tab:obr-inputs-outputs} shows example input-output pairs for Optical Braille Recognition.
\begin{table*}[tb]
     \small
     \centering
     \renewcommand{\arraystretch}{1.3}
     \begin{tabular}{@{}l l@{}}
     \toprule
     {\includegraphics[width=0.4\textwidth]{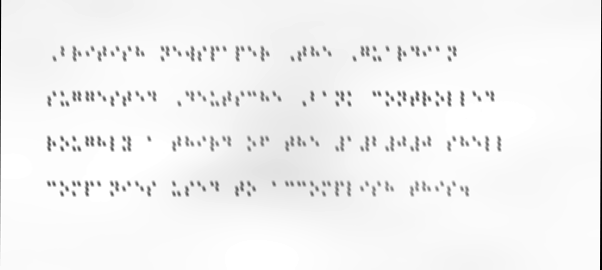}} & \raisebox{1.7\height}{\begin{tabular}{@{}l@{}}\textbf{Source: WMT2024}\\\textbf{Input:} Transcribe the Braille to English.\\\textbf{Output:} \colorbox{orange!10!beige}{British newspaper The Guardian suggested Deutsche Bank controlled} \\\colorbox{orange!10!beige}{roughly a third of the 1200 shell companies used to accomplish this.}\end{tabular}}\\
     \midrule
    {\includegraphics[width=0.4\textwidth]{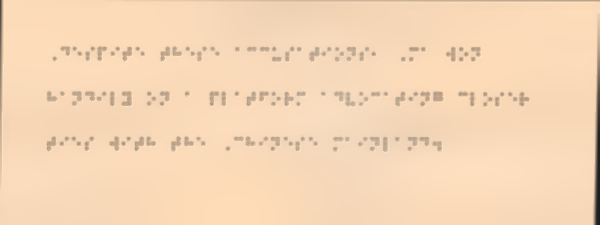}} & \raisebox{1.3\height}{\begin{tabular}{@{}l@{}}\textbf{Source: WMT2024}\\\textbf{Input:} Transcribe the Braille to English.\\\textbf{Output:} \colorbox{orange!10!beige}{Despite these accusations, Ma won handily on a platform advocating} \\\colorbox{orange!10!beige}{closer ties with the Chinese mainland.}\end{tabular}}\\
    \midrule
    {\includegraphics[width=0.4\textwidth]{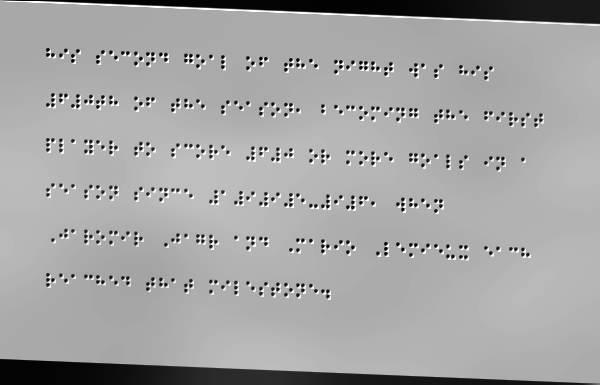}} & \raisebox{2.2\height}{\begin{tabular}{@{}l@{}}\textbf{Source: WMT2024}\\\textbf{Input:} Transcribe the Braille to English.\\\textbf{Output:} \colorbox{orange!10!beige}{his second goal of the night was his 60th of the season, becoming} \\\colorbox{orange!10!beige}{the first player to score 60 or more goals in a season since 1995-96,}\\\colorbox{orange!10!beige}{when Jaromir Jagr and Mario Lemieux each reached that milestone.}\end{tabular}}\\
    \midrule
    {\includegraphics[width=0.4\textwidth]{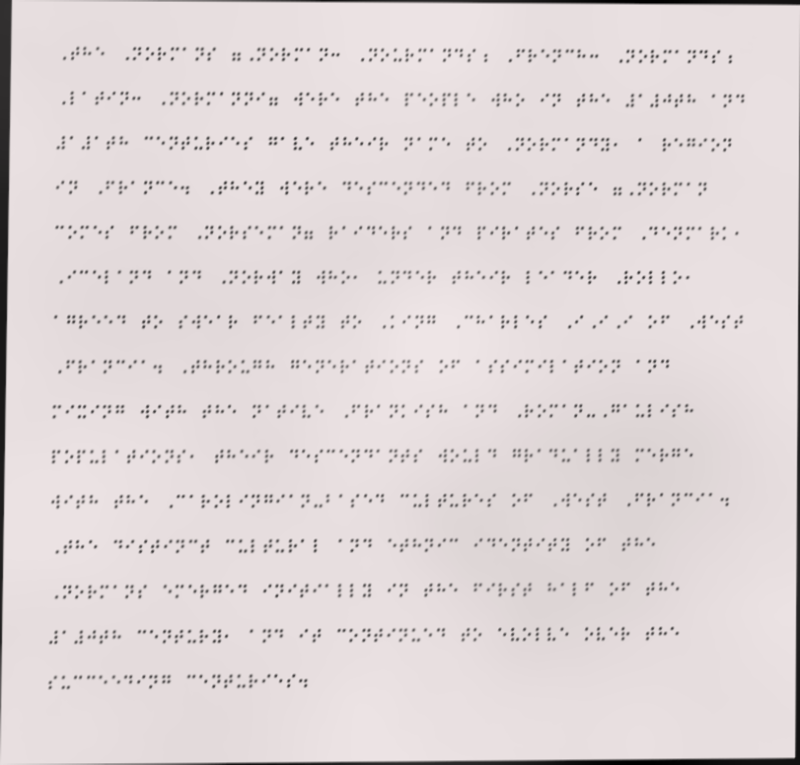}} & \raisebox{3.9\height}{\begin{tabular}{@{}l@{}}\textbf{Source: SquAD}\\\textbf{Input:} Answer the following question based on the image.\\If the question is not answerable, output `unanswerable'.\\When were the Normans in Normandy?\\\textbf{Output:} \colorbox{orange!10!beige}{10th and 11th centuries}\end{tabular}}\\
    \midrule
    {\includegraphics[width=0.4\textwidth]{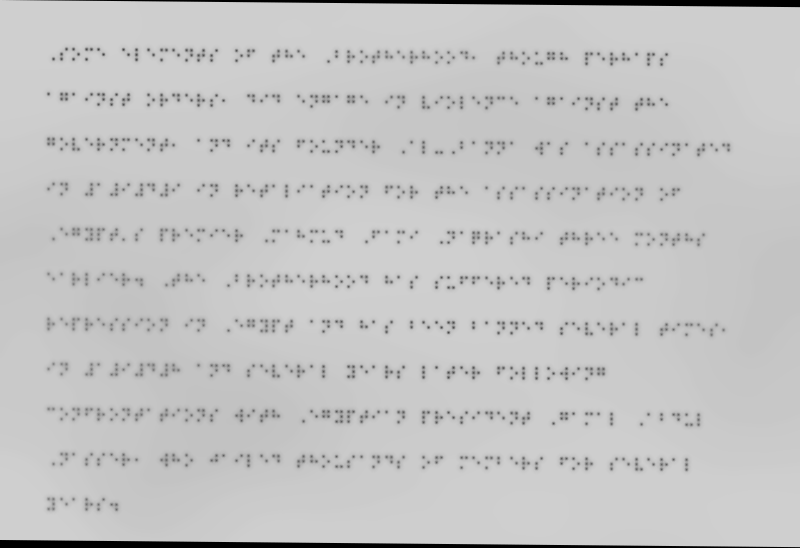}} & \raisebox{2.55\height}{\begin{tabular}{@{}l@{}}\textbf{Source: SquAD}\\\textbf{Input:} Answer the following question based on the image.\\If the question is not answerable, output `unanswerable'. \\What Egyptian president jailed hundreds of members of the Brotherhood?\\\textbf{Output:} \colorbox{orange!10!beige}{unanswerable}\end{tabular}}\\

      \bottomrule
      \end{tabular}
      \caption{Illustration of inputs-outputs for the two Optical Braille Recognition tasks: Braille-to-Text Transcription and Cross-Script Question Answering. \colorbox{orange!10!beige}{Highlighted text} indicates the target output. Note that the model cannot provide the correct response unless it can map Braille symbols to English.}
      \label{tab:obr-inputs-outputs}
  \end{table*}

\section{Acknowledgments}
We acknowledge the use of GitHub Copilot\footnote{\href{https://github.com/features/copilot}{https://github.com/features/copilot}} in the implementation of our research. 
All final code is verified by the authors.

\end{document}